\documentclass[fleqn,usenatbib,usedcolumn]{mnras}

\usepackage[T1]{fontenc}

\DeclareRobustCommand{\VAN}[3]{#2}
\let\VANthebibliography\thebibliography
\def\thebibliography{\DeclareRobustCommand{\VAN}[3]{##3}\VANthebibliography}

\usepackage[british]{babel}
\usepackage{graphicx}	
\usepackage{amsmath}	
\usepackage{amssymb}	
\usepackage{multirow}
\usepackage{tabularx}

\usepackage[inline]{enumitem}
\usepackage{newtxtext}
\usepackage[slantedGreek]{newtxmath}
\usepackage{placeins}

\usepackage[inline]{enumitem}
\usepackage{soul} 
\usepackage{amsmath}
\usepackage{newtxtext}
\usepackage[slantedGreek]{newtxmath}
\usepackage{xifthen}
\usepackage{xspace}

\usepackage[hyperref]{ntheorem}
\newtheorem{quest}{Question}[section]

\newtheorem{disc}{Discussion}[section]

\newcommand{\orcid}[1]{\href{https://orcid.org/#1}{\,\includegraphics[height=\fontcharht\font`\B]{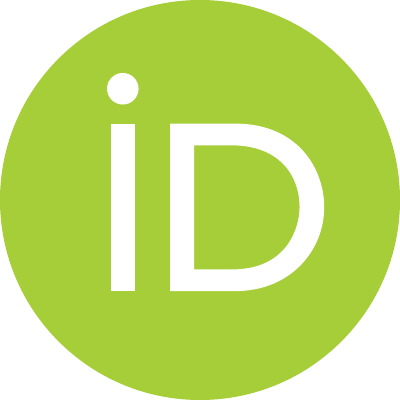}}}
\newcommand{\github}[1]{\href{https://github.com/#1}{\includegraphics[height=\fontcharht\font`\B]{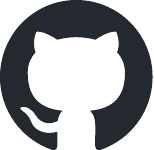} \nolinkurl{#1}}}

\newcommand{\aperture}{\ckpc[30]}

\newcommand{\percent}[1]{\ensuremath{#1}~per cent}

\newcolumntype{L}{>{\raggedright\arraybackslash}X}
\newcolumntype{Y}{>{\centering\arraybackslash}X}
\newcolumntype{R}{>{\raggedleft\arraybackslash}X}

\newcommand{\PLANCKfifteenGMEAGLEp}{\ensuremath{{\Ho{}=\kms[67.32]\pMpc{}},\,\allowbreak {\Omegaup_{\rm m}=0.3156},\,\allowbreak {\Omegaup_\Lambdaup=0.6844},\,\allowbreak {\Omegaup_{\rm b}=0.0498},\,\allowbreak {n_{\rm s}=0.9645},\,\allowbreak {\sigma_8=0.8310}}}

\makeatletter
\newcommand{\phantomlabel}[2]{
    \protected@write\@auxout{}{
        \string\newlabel{#2}{
            {\@currentlabel~(#1)}{\thepage}
            {\@currentlabel~(#1)}{#2}{}
        }
    }%
    \hypertarget{#2}{}%
}\newcommand{\alphantomlabel}[2]{
    \protected@write\@auxout{}{
        \string\newlabel{#2}{
            {\@currentlabel,~#1}{\thepage}
            {\@currentlabel,~#1}{#2}{}
        }
    }%
    \hypertarget{#2}{}%
}
\makeatother

\newcommand{\secref}[1]{Section~\ref{#1}}

\newcommand{\figref}[1]{Fig.~\ref{#1}}
\newcommand{\figrefs}[2]{Figs~\ref{#1}~and~\ref{#2}}
\newcommand{\eqnref}[1]{eq.~(\ref{#1})}

\newcommand{\extfig}[1]{fig.~\ensuremath{#1}}

\newcommand{\tabref}[1]{Table~\ref{#1}}

\newcommand{\Gaia}{{\textit{Gaia}}}

\newcommand{\Eagle}{{EAGLE}}

\newcommand{\Suppressed}{{\sc suppressed}}
\newcommand{\Organic}{{\sc organic}}
\newcommand{\Enhanced}{{\sc enhanced}}

\newcommand{\Gadget}{{\sc gadget}}

\newcommand{\GadgetIII}{\Gadget{}3}
\newcommand{\Genetic}{{\sc genetIC}}

\newcommand{\PGadgetIII}{{\sc p-}\GadgetIII{}}
\newcommand{\astropy}{{\sc astropy}}
\newcommand{\emosaics}{E-MOSAICS}
\newcommand{\mosaics}{{\sc mosaics}}
\newcommand{\matplotlib}{{\sc matplotlib}}
\newcommand{\numpy}{{\sc numpy}}
\newcommand{\pynbody}{{\sc pynbody}}
\newcommand{\python}{{\sc python}}
\newcommand{\scipy}{{\sc scipy}}
\newcommand{\subfind}{{\sc subfind}}
\newcommand{\tangos}{{\sc tangos}}

\newcommand{\Fhalo}{\ensuremath{F^{\rm halo}}}

\newcommand{\gc}{GC}
\newcommand{\ics}{initial conditions}
\newcommand{\Jcirc}{\ensuremath{J_{\rm circ}}}

\newcommand{\Jz}{\ensuremath{J_{\rm z}}}
\newcommand{\LCDM}{{$\Lambdaup$CDM}}
\newcommand{\M}[1]{\ensuremath{M_{#1}}}
\newcommand{\m}[1]{\ensuremath{m_{#1}}}

\newcommand{\Mgc}{\ensuremath{\M{\rm GC}}}
\newcommand{\Nbody}{\mbox{\textit{N}--body}}
\newcommand{\Ngc}{\ensuremath{N_{\rm GC}}}
\newcommand{\R}[1]{\ensuremath{R_{#1}}}

\newcommand{\Rvir}{\ensuremath{\R{200}}}
\newcommand{\SMBH}{SMBH}

\newcommand{\unit}[1]{\ensuremath{\mathrm{\,#1}}\xspace}
\newcommand{\unitlogicnospace}[2]{%
  \ifthenelse{\isempty{#1}}%
    {\unit{#2}}
    {\ensuremath{{{#1}\unit{#2}}}}
  }
\newcommand{\unitlogicspace}[2]{%
  \ifthenelse{\isempty{#1}}%
    {\unit{#2}}
    {\ensuremath{{{#1}\, \unit{#2}}}}
  }

\newcommand{\Msun}[1][]{\unitlogicspace{#1}{M_\odot}}
\newcommand{\Msunh}[1][]{\unitlogicspace{#1}{\mathit{h}^{-1}\Msun{}}}
\newcommand{\Msunyr}[1][]{\unitlogicspace{#1}{M_\odot\, {\rm yr^{-1}}}}
\newcommand{\Mpc}[1][]{\unitlogicspace{#1}{Mpc}}

\newcommand{\kpc}[1][]{\unitlogicspace{#1}{kpc}}
\newcommand{\ckpc}[1][]{\unitlogicspace{#1}{ckpc}}
\newcommand{\pc}[1][]{\unitlogicspace{#1}{pc}}

\newcommand{\pMpc}[1][]{\unitlogicspace{#1}{Mpc^{-1}}}

\newcommand{\Gyr}[1][]{\unitlogicspace{#1}{Gyr}}

\newcommand{\kms}[1][]{\unitlogicspace{#1}{km\, s^{-1}}}
\newcommand{\KcmCubed}[1][]{\unitlogicspace{#1}{K\, cm^{-3}}}

\newcommand{\variablelogicspace}[2]{%
  \ifthenelse{\isempty{#2}}%
    {\ensuremath{#1}}
    {\ensuremath{{{#1}={#2}}}}
  }

\newcommand{\kcostar}[1][]{\variablelogicspace{\kappa_{\rm co,\, \ast}}{#1}}

\newcommand{\tlb}[1][]{\variablelogicspace{t_{\rm lookback}}{#1}}
\newcommand{\z}[1][]{\variablelogicspace{z}{#1}}
\newcommand{\CFE}[1][]{\variablelogicspace{\Gammaup}{#1}}
\newcommand{\Ho}[1][]{\variablelogicspace{H_0}{#1}}
\newcommand{\Mhalo}[1][]{\variablelogicspace{\M{200}}{#1}}
\newcommand{\Lstar}[1][]{\variablelogicspace{L^{\ast}}{#1}}

\newcommand{\Mclust}[1][]{\variablelogicspace{\M{\rm cl}}{#1}}
\newcommand{\Mclusti}[2][]{\variablelogicspace{\M{{\rm cl,}\, i}\!\left({#1}\right)}{#2}}
\newcommand{\Mcstar}[1][]{\variablelogicspace{\M{\rm c,\, \ast}}{#1}}
\newcommand{\MgasSF}[1][]{\variablelogicspace{\M{\rm gas,\, SF}}{#1}}

\newcommand{\Msmbh}[1][]{\variablelogicspace{\M{\rm SMBH}}{#1}}
\newcommand{\Mstar}[1][]{\variablelogicspace{\M{\ast}}{#1}}
\newcommand{\Mstarh}[1][]{\variablelogicspace{\M{\ast}^{\rm halo}}{#1}}
\newcommand{\Mstari}[1][]{\variablelogicspace{\M{\ast,\, i}}{#1}}

\newcommand{\SM}[1][]{\variablelogicspace{S_{\rm M}}{#1}}
\newcommand{\TN}[1][]{\variablelogicspace{T_{\rm N}}{#1}}

\graphicspath{{Figures/}}

\title[Globular cluster formation \& evolution in controlled galaxy simulations]{The formation and disruption of globular cluster populations in simulations of present-day \Lstar{} galaxies with controlled assembly histories}

\author[O. Newton et al.]{Oliver Newton\orcid{0000-0002-2769-9507}\thanks{E-mail: onewton@cft.edu.pl},$^{1,2}$
Jonathan J. Davies\orcid{0000-0002-8337-3659},$^{1,3}$
Joel Pfeffer\orcid{0000-0003-3786-8818},$^{4}$
Robert A. Crain\orcid{0000-0001-6258-0344},$^{1}$
\newauthor J.~M.~Diederik Kruijssen\orcid{0000-0002-8804-0212},$^{5,6}$
Andrew Pontzen,$^{7}$
and Nate Bastian$^{8,9}$
\\
$^{1}$Astrophysics Research Institute, Liverpool John Moores University, 146 Brownlow Hill, Liverpool L3 5RF, UK\\
$^{2}$Centre for Theoretical Physics, Polish Academy of Sciences, al. Lotnik\'{o}w 32/46 Warsaw, Poland\\
$^{3}$Department of Physics and Astronomy, University College London, Gower Street, London WC1E 6BT, UK\\
$^{4}$Centre for Astrophysics \& Supercomputing, Swinburne University, Hawthorn, VIC 3122, Australia\\
$^{5}$Technical University of Munich, School of Engineering and Design, Department of Aerospace and Geodesy, Chair of Remote Sensing Technology, \\\hspace{2.2mm}Arcisstr. 21, 80333 Munich, Germany\\
$^{6}$Cosmic Origins Of Life (COOL) Research DAO, coolresearch.io\\
$^{7}$Institute for Computational Cosmology, Durham University, South Road, Durham, DH1 3LE, UK\\
$^{8}$Donostia International Physics Center (DIPC), Paseo Manuel de Lardizabal, 4, E-20018 Donostia-San Sebasti\'an, Guipuzkoa, Spain\\
$^{9}$IKERBASQUE, Basque Foundation for Science, E-48013 Bilbao, Spain%
\vspace{-10pt}%
}
\date{Accepted 2025~July~15. Received 2025~July~10; in original form 2024~September~20}

\pubyear{2025}

\begin{document}
\label{firstpage}
\pagerange{\pageref{firstpage}--\pageref{lastpage}}
\maketitle

\begin{abstract}
Globular clusters (GCs) are sensitive tracers of galaxy assembly histories but interpreting the information they encode is challenging because mergers are thought to promote both the formation and disruption of \gc{}s. We use simulations with controlled merger histories to examine the influence of merger mass ratio on the \gc{} population of a present-day \Lstar{} galaxy, using the genetic modification technique to adjust the initial conditions of a galaxy that experiences major mergers at \z[1.7] and 0.77 (\Organic{} case), so the later merger has twice its original mass ratio (\Enhanced{} case), or is prevented from occurring (\Suppressed{} case). We evolve the three realizations with \emosaics{} (MOdelling Star cluster population Assembly In Cosmological Simulations with \Eagle{}), which couples subgrid star cluster formation and evolution models to the \Eagle{} (Evolution and assembly of GaLaxies and their Environments) galaxy formation model. Relative to the \Organic{} case, the mass of surviving \gc{}s is elevated (reduced) in the \Enhanced{} (\Suppressed{}) case, indicating that major mergers promote a net boost to the \gc{} population. The boost is clearly quantified by the \gc{} specific mass, $S_{\rm M}$, because it is sensitive to the number of the most massive \gc{}s, whose long characteristic disruption time-scales enable them to survive their hostile natal environments. In contrast, the specific frequency, $T_{\rm N}$, is insensitive to assembly history because it primarily traces low-mass \gc{}s that tend to be disrupted soon after their formation. The promotion of \gc{} formation and disruption by major mergers imprints a lasting and potentially observable signature: an elevated mass fraction of field stars in the galaxy's stellar halo that were born in star clusters. 
\end{abstract}

\begin{keywords}
Galaxy: evolution -- globular clusters: general -- galaxies: evolution -- galaxies: interactions -- galaxies: star clusters: general -- galaxies: star formation%
\vspace{-20pt}%
\end{keywords}



\section{Introduction}
\label{sec:Introduction}
In the prevailing cosmological paradigm dark matter haloes grow by smooth accretion and mergers with other haloes \citep{white_core_1978,blumenthal_formation_1984}. The gas accreted in tandem with the dark matter dissipates its gravitational potential energy via radiative cooling, and seeds the formation of galaxies. Their subsequent growth is driven by further gas accretion, and mergers with other galaxies. These mechanisms, in combination with regulatory feedback processes, drive the evolution of the galaxy population \citep[e.g.][]{white_galaxy_1991}.

Numerical simulations of the formation and evolution of galaxies incorporating these processes have attained a degree of maturity such that they reproduce a diverse range of properties of the observed galaxy population \citep[for a recent review see][]{crain_hydrodynamical_2023}. Such simulations indicate that although mergers between galaxies of similar mass occur infrequently, their effect on the subsequent evolution of the remnant galaxy can be profound. Major mergers such as these are expected to drive a large fraction of the interstellar medium towards the inner regions of galaxies, triggering intense star formation and accelerating the growth of the central supermassive black hole \citep[SMBH; e.g.][]{heckman_galaxy_1986,hernquist_tidal_1989,barnes_transformations_1996,mihos_gasdynamics_1996,springel_modelling_2005,cox_effect_2008,davies_are_2024}. Moreover, these events significantly reorganize the internal structure of the remnant galaxies, leaving long-lived signatures on their baryonic components that potentially remain observable today \citep{fakhouri_nearly_2008,stewart_galaxy_2009,wetzel_simulating_2009,hoffman_orbital_2010,lagos_quantifying_2018,pulsoni_stellar_2021,sotillo-ramos_merger_2022,byrne-mamahit_interacting_2023,santucci_distribution_2024}.

The extreme physical conditions that arise during major mergers can markedly influence the evolution of galaxies and encode detailed information about the progenitor galaxies within the remnant. Ideally, this would be probed using resolved studies of individual stars; however, at present such investigations are extremely challenging to carry out in all but the closest galaxies \citep[e.g.][]{newton_undiscovered_2023}. An alternative approach to trace the rich assembly histories of galaxies entails using globular clusters~(GCs). These massive, compact, old, gravitationally bound systems of stars exhibit high surface brightnesses, and hence are observable at large distances (\citealp{harris_catalog_1996,marin-franch_acs_2009,carretta_properties_2010,harris_catalog_2013,vandenberg_ages_2013,baumgardt_mean_2019}, \citealp[also see e.g.][for reviews]{harris_globular_1979,harris_globular_1991,brodie_extragalactic_2006}). These properties may facilitate access to the formation histories of a statistical sample of galaxies with which to test models of structure formation and further refine models of galaxy formation. To take advantage of this the mechanisms governing the formation and evolution of the \gc{}s themselves must be understood.

Contemporary models of star cluster formation posit that the most massive clusters form within dense and highly pressurized gas in the interstellar medium \citep[][see also \citealp{krumholz_star_2019}; \citealp{kruijssen_formation_2025} for recent reviews]{mckee_formation_2003,kravtsov_formation_2005,kruijssen_globular_2014,pfeffer_e-mosaics_2018,keller_where_2020}. Among others, this picture has been motivated by observations (see below), theory \citep[e.g.][]{elmegreen_universal_1997} and numerical simulations, including low-resolution models of spiral galaxies \citep[e.g.][]{bekki_globular_2002,li_formation_2004}, and high-resolution simulations of resolved cluster formation in merging dwarf or high-redshift galaxies \citep[e.g.][]{lahen_formation_2019,ma_self-consistent_2020}. Collectively, simulations show that high gas pressures and densities promote the formation of massive clusters \citep{kruijssen_formation_2012,lahen_formation_2019,lahen_griffin_2020,brown_testing_2022,sameie_formation_2023,joschko_cosmic_2024}. Similar results have also been demonstrated in simulations of major mergers involving galaxies with similar masses to the Milky Way that are found to promote the formation of massive star clusters \citep{li_star_2017,li_formation_2022}.

These conditions, which are extreme in the present-day cosmos, were more prevalent at early epochs and likely promoted the formation of \gc{}s in the progenitors of present-day massive galaxies \citep{kruijssen_formation_2025}. Observation-based modelling of \gc{} ages suggests that most of them formed close to the peak of star formation in their host galaxies \citep[e.g.][]{kerber_physical_2007,dias_age_2010,dotter_acs_2010,dotter_globular_2011,caldwell_star_2011,beasley_evidence_2015,usher_sluggs_2019}. At later times, similarly favourable conditions for massive cluster formation emerge during significant galaxy mergers. The most massive star clusters forming today all reside within galaxies experiencing gas-rich major mergers \citep{holtzman_planetary_1992,schweizer_ages_1998,whitmore_luminosity_1999}. The strong correlation between \gc{} formation rates and the availability of plentiful supplies of cold gas implies that such merger events exert considerable influence on the distributions of age and metallicity of the \gc{} system \citep{cote_formation_1998,kissler-patig_constraints_1998,larsen_young_2000}. Consequently, galaxy mergers have been the focus of other studies using semi-analytic \citep[e.g.][]{muratov_modeling_2010,li_modeling_2014,li_star_2017,choksi_formation_2018} and numerical simulations \citep[e.g.][]{ricotti_common_2016,renaud_origin_2017,creasey_globular_2018,el-badry_formation_2019,lahen_griffin_2020,ma_self-consistent_2020,valenzuela_globular_2021,chen_catalogue_2024} to probe their effect on the formation and demographics of \gc{} systems.

The stars contributed to the stellar halo by disrupted \gc{}s are particularly valuable to reconstruct galaxy assembly histories because they are chemically distinct from normal halo field stars \citep{martell_light-element_2010,martell_chemical_2016,schiavon_chemical_2017,horta_contribution_2021}; many of the latter originate from the accretion of stellar and gaseous material stripped from low-mass galaxies during minor mergers \citep[][but see also \citealt{font_cosmological_2011}]{searle_composition_1978,forbes_origin_1997,cote_formation_1998,cote_evidence_2000,bullock_tracing_2005}. Indeed, this has been demonstrated especially lucidly in recent years by exquisite observational measurements of \gc{}s in the Milky Way using \Gaia{}. Kinematic analyses of these data suggest that a significant fraction of halo \gc{}s in the Milky Way likely originated from the merger with the Gaia--Enceladus galaxy \citep{belokurov_co-formation_2018,helmi_merger_2018,myeong_sausage_2018,massari_origin_2019}. Furthermore, several other galaxy mergers have been identified as contributors to the build-up of the stellar disc and the stellar halo \citep{ibata_dwarf_1994,helmi_debris_1999,kruijssen_formation_2019,myeong_evidence_2019,kruijssen_kraken_2020,horta_linking_2021}. Recent work suggests that \percent{{35-50}} of the Milky Way \gc{} population may have formed or been accreted during such mergers \citep[e.g.][]{kruijssen_formation_2019,massari_origin_2019,kruijssen_kraken_2020}; however, associating \gc{}s with a progenitor galaxy is difficult and would benefit from more precise age estimates for confirmation \citep{helmi_streams_2020}.

Typically, studies that examine the connection between the assembly histories of galaxies and their present-day properties are carried out using ensembles of systems extracted from hydrodynamical simulations of large cosmological volumes. Such approaches have identified correlations between the growth and evolution of galaxies and various non-linear baryonic processes, such as star formation, radiative feedback, and other gas-dynamical interactions \citep{matthee_origin_2017,davies_gas_2019,davies_quenching_2020,montero-dorta_influence_2021}. Despite this success, it is challenging to establish an unambiguous causal relationship between these processes and the effects that arise due to the assembly histories of the galaxies because each galaxy in the sample is situated in a different environment. The properties of haloes correlate strongly with their cosmic environments \citep{sheth_environmental_2004,gao_age_2005}, so this must be controlled for when studying how the assembly history of a galaxy affects its evolution.

Studying ensembles of galaxies drawn from simulations is also limited by the small sample sizes within a given environment. Simulating (at high resolution) volumes that are large enough to capture significant diversity in the assembly histories of the galaxies would be computationally prohibitive. Using large volume zoom-in simulations partially addresses these limitations \citep[e.g.][]{crain_galaxies-intergalactic_2009,lovell_first_2021,newton_hermeian_2022-1}; however, the initial conditions are complicated to devise, and the simulations still fail to fully disentangle the correlation between the assembly history and the cosmic environment. In our study we address this by adopting the `genetic modification' technique \citep{roth_genetically_2016,pontzen_how_2017}, in which the initial conditions governing the formation of a target galaxy are carefully modified to change targeted aspects of its assembly history, whilst leaving the large-scale environment of the galaxy unchanged. Controlled experiments such as these isolate the effects of small variations in the assembly history from the response to the nearby environment, thereby minimizing other confounding influences \citep[see also][]{davies_quenching_2021,davies_galaxy_2022,davies_are_2024}.

In this paper, we introduce a new suite of cosmological hydrodynamic zoom-in simulations of a present-day \Lstar{} galaxy that we use to characterize how the mass ratio of major mergers influences the formation and disruption of \gc{}s and their contribution to the build-up of stellar haloes. For the first time we combine two state-of-the-art techniques to facilitate a novel study of the causal relationship between galaxy assembly history and the evolution of the \gc{} population. First, while previous work has addressed the influence of assembly history by studying an ensemble of galaxies drawn from diverse cosmic environments, here we adopt the complementary approach of focusing on a single galaxy and modifying its \ics{} to perform a controlled experiment in which the assembly history is adjusted. This technique dissociates the effect of small changes to the assembly history from those arising from confounding environmental influences. Second, the cluster formation model we adopt accounts for the physical properties of the natal environment during formation, which can affect the survivability of clusters shortly after their birth. It links the evolution and disruption of star clusters to the properties of the local tidal field in the immediate vicinity of the cluster, and naturally accounts for changes in the global galaxy properties in response to merger activity.

We organize this paper as follows. \secref{sec:Methods} introduces the physical models and numerical simulations that we use. We also describe the genetic modification technique and the alterations that we make to the galaxy's assembly history. We present our main results in \secref{sec:Results}, detailing how the \gc{} population responds to the changes we make to the \ics{} and the contribution of disrupted clusters to the build-up of the stellar halo. We summarize our results and contextualize them with other theoretical and observational studies in \secref{sec:Sum_Disc}.

\section{Methods}
\label{sec:Methods}
The simulations were evolved using a modified version of the \Nbody{} \textsc{TreePM} smoothed particle hydrodynamics (SPH) \PGadgetIII{} code \citep[last described by][]{springel_simulations_2005} to solve the coupled equations of gravity and hydrodynamics. This version of \PGadgetIII{} was developed for the \Eagle{} (Evolution and assembly of GaLaxies and their Environments) simulations of galaxy formation and features several changes to the numerical algorithms and the subgrid models. Included in the latter are the treatments of star cluster formation and disruption developed for the \emosaics{} (MOdelling Star cluster population Assembly In Cosmological Simulations with \Eagle{}) project. The \Eagle{} and \emosaics{} models are described in many other publications, so we restrict ourselves to brief descriptions of their key components and direct interested readers to the reference papers of the respective projects.

\subsection{The \Eagle{} galaxy formation model}
\label{sec:Methods:Eagle}
The \Eagle{} project \citep{schaye_eagle_2015,crain_eagle_2015} is a suite of hydrodynamical simulations of galaxy formation in the Lambda-cold dark matter~(\LCDM{}) cosmogony, whose raw data and processed data products have been released to the community \citep{mcalpine_eagle_2016}. The \Eagle{} model includes modifications to the standard \PGadgetIII{} SPH implementation and time-stepping criteria, and a suite of subgrid models to govern processes acting on scales below the numerical resolution limit of the simulation. Of particular relevance for this work are the photo-heating and radiative cooling of $11$ individual chemical elements that contribute significantly to the cooling of photoionized plasmas \citep{wiersma_effect_2009}; the seeding of \SMBH{}s at the centres of haloes with masses greater than \Msunh[{10^{10}}], and the associated stochastic thermal feedback \citep{booth_cosmological_2009,schaye_eagle_2015}; and the stochastic formation of star particles \citep{schaye_relation_2008} from gas particles with densities greater than a metallicity dependent threshold \citep{schaye_star_2004}. Each star particle is modelled as a simple stellar population with a \citet{chabrier_galactic_2003} initial mass function. The evolution of stellar populations and the associated stellar mass loss is governed by the \citet{wiersma_chemical_2009} model. This accounts for the variation in stellar lifetimes as a function of mass and metallicity \citep{portinari_galactic_1998} and the chemical enrichment of the interstellar medium by Type~Ia and Type~II supernovae, and stars experiencing the asymptotic giant branch phase. The rate of Type~Ia supernovae is modelled with an empirical exponential delay time distribution function.

\subsection{E-MOSAICS}
\label{sec:Methods:Emosaics}
\emosaics{} \citep{pfeffer_e-mosaics_2018,kruijssen_e-mosaics_2019} is a suite of \Eagle{} spin-off simulations that follow the co-formation and co-evolution of galaxies and their star cluster populations. For these simulations, modified versions of the \mosaics{} semi-analytic models of star cluster formation and evolution \citep{kruijssen_modelling_2011} were added to the \Eagle{} version of \PGadgetIII{} as subgrid models. This implementation describes the formation and evolution of subgrid populations of bound star clusters that are `attached' to the stellar particles that form in the \Eagle{} model \citep{pfeffer_e-mosaics_2018,kruijssen_e-mosaics_2019}. In this model \gc{}s represent the extreme high-gas pressure and high-density tail of star cluster formation.

The birth properties of star clusters are determined by those of their natal gas particle at the time of formation. The initial fraction of the stellar particle's mass composed of star clusters is termed the `cluster formation efficiency', \CFE{} \citep{bastian_star_2008}, and in the \citet{kruijssen_fraction_2012} model that we adopt \CFE{} is a strong function of the local gas pressure. The initial masses of the clusters are drawn stochastically from a cluster birth mass function with a \citet{schechter_analytic_1976} functional form that is truncated beyond an upper mass scale, \Mcstar{}. The value of \Mcstar{} increases as a function of the natal gas pressure and density but decreases in regions of high shear and centripetal forces, such as in galaxy centres \citep{reina-campos_unified_2017}. This approach to populating the subgrid cluster population means that some stellar particles will carry no star clusters, while a small fraction of the remaining stellar particles may carry more mass in clusters than the dynamical mass of the host particle. However, on average the fraction of the newly formed stellar mass in clusters will be equal to \CFE{}, as required. Clusters that form with initial masses less than \Msun[{5\times10^3}] are discarded because they disrupt on short time-scales. For simplicity we assume that star clusters are born with a half-mass radius of \pc[4] and omit the influence of cluster radius evolution.

Cluster masses evolve via stellar mass loss, which is tracked self-consistently by the \Eagle{} chemodynamics implementation discussed in \secref{sec:Methods:Eagle}, and dynamically in response to the local environment of the host stellar particle. Whereas stellar evolution transfers mass from stars (irrespective of whether they are field stars or in clusters) to the gas phase, dynamical mass loss transfers mass from the clustered component of the stellar particles to the field component. Dynamical mass loss proceeds via tidal shocks and two-body relaxation according to the models of \citet{kruijssen_modelling_2011}. Both mechanisms are governed by the local tidal field and the instantaneous mass of the cluster, and are computed on-the-fly at each time step of the simulation. Star clusters are also subject to the effects of dynamical friction, which cannot be computed self-consistently because each stellar particle hosts clusters with a range of masses and associated dynamical friction time-scales. Therefore, we implement an approximate treatment as a post-processing step applied sequentially to all snapshots in which any cluster older than its dynamical friction time-scale, computed using the expression of \citet{lacey_merger_1993}, is disrupted manually. All clusters are tracked until they reach a minimum cluster mass of \Mclust[{\Msun[10^2]}], below which they are considered fully disrupted.

Studies using the \emosaics{} simulations have adopted various definitions of the clusters that are considered \gc{}s to facilitate the most appropriate comparison with observational data. For example, using thresholds in cluster mass and age \citep[e.g.][]{reina-campos_dynamical_2018,usher_origin_2018,kruijssen_e-mosaics_2019,hughes_fefeh_2020,keller_where_2020,pfeffer_predicting_2020,reina-campos_constraining_2023}, or adopting a galaxy mass-dependent cluster mass threshold \citep[e.g.][]{bastian_globular_2020,reina-campos_radial_2022,pfeffer_globular_2023,trujillo-gomez_situ_2023}. In this work we examine the trends in the simulations rather than making detailed comparisons with observational data, so we simplify the interpretation of our results by defining \gc{}s as star clusters with masses, $\Mclust{} \geq \Msun[10^5]$ \citep[similar to e.g.][]{hughes_fossil_2019,horta_linking_2021,shao_survival_2021,trujillo-gomez_kinematics_2021,dolfi_present-day_2022}.

\subsection{Target selection and construction of zoom initial conditions}
\label{sec:Methods:Target}
We begin by simulating a periodic volume of the \LCDM{} cosmogony adopting parameters derived from the \citet{ade_planck_2016} data (\PLANCKfifteenGMEAGLEp{}), whose \ics{} were generated at \z[99] using \Genetic{} \citep{stopyra_geneticnew_2021}. The cubic volume has sides of length $L=\Mpc[50]$, and was realized with $512^3$ dark matter particles of mass, $\m{\rm DM}=\Msun[{3.2\times10^7}]$, and an equal number of gas particles with mass, $\m{\rm gas}=\Msun[{5.6\times10^6}]$. The volume was evolved to the present day with the \Eagle{} model described in \secref{sec:Methods:Eagle} adopting the `Reference' parameters \citep[see][]{schaye_eagle_2015}.  

From this volume we select a `typical' present-day \Lstar{} galaxy with stellar mass $\Mstar[{\Msun[{4.5\times10^{10}}]}]$, and dark matter halo mass $\Mhalo[{\Msun[{3.3\times10^{12}}]}]$. Its assembly history includes two significant mergers: a major merger at \z[1.7] in which the ratio of stellar masses of the infalling galaxy to the galaxy of interest is $\mu_\ast = 0.45$, followed by a less significant and final merger at \z[0.77] with $\mu_\ast = 0.16$. The galaxy lies on the present-day star formation main sequence and has a circumgalactic medium gas mass fraction that is typical for a galaxy of this mass realized in the \Eagle{} simulations \citep[see e.g.][]{davies_gas_2019}. It is also isolated, such that there are no dark matter haloes of greater mass within \Mpc[3.3]; a requirement imposed to minimize contamination due to tidal effects. The full set of selection criteria that we apply is described in detail by \citet{davies_quenching_2021}. 

We generate `zoom-in' \ics{} for this galaxy by selecting all particles within three virial radii of the galaxy at \z[0] and identifying the Lagrangian region defined by these particles in the \ics{} of the parent volume. We refine this region with a factor of $3^3=27$ more particles, yielding particle masses of $\m{\rm DM}=\Msun[1.18\times10^6]$ and $\m{\rm gas}=\Msun[2.17\times10^5]$. These are comparable to those of the high-resolution \Eagle{} volumes and the \emosaics{} simulations. We also downsample the mass distribution external to the refinement region by a factor of $2^3=8$, yielding low-resolution boundary particles of mass $\m{\rm DM,\, LR}=\Msun[3.02\times10^8]$. We refer to the galaxy that forms from these \ics{} as the \Organic{} case.

At \z[0.77] the \Organic{} galaxy experiences a significant merger with $\mu_\ast = 0.16$. Previous studies examining this simulated galaxy found that altering the mass ratio of this merger can induce significant changes to the galaxy’s star formation history, kinematics, and circumgalactic medium \citep{davies_galaxy_2022}. In this work we investigate the influence of the merger mass ratio on the properties of the galaxy's \gc{} population. Following the procedure described by \citet{davies_galaxy_2022}, we use the genetic modification technique to generate two sets of modified \ics{} in which the significance of the final merger is altered. In the \Enhanced{} case the merger takes place at \z[0.89] and has a stellar mass ratio of $\mu_\ast=0.36$. The incoming galaxy is twice as massive and the merger takes place earlier than that experienced by the \Organic{} system during its assembly. In the \Suppressed{} set of \ics{} the significance of the final merger is reduced so severely that it does not take place at all. The mass ratio of an eventual merger taking place after \z[0] would be negligible. We simulate the same galaxy as \citet{davies_galaxy_2022}, but note that some of the baryonic properties of the galaxy presented here differ slightly from those presented in the earlier study. This is because both studies adopt the subgrid parameters of the \Eagle{} `Recal' model (see Section \ref{sec:Methods:Simulations}) but here we adopt a higher resolution (particle masses lower by a factor $\simeq 3.4$) than \citet{davies_galaxy_2022}, which is slightly closer to that of the high-resolution \Eagle{} simulations used to calibrate the subgrid parameters of the `Recal' model. Consequently, there are differences stemming from the imperfect `strong convergence' behaviour of the model \citep[for further discussion of this issue see section 2 of][]{schaye_eagle_2015}.

\subsection{Simulations}
\label{sec:Methods:Simulations}

We evolve the \ics{} of the \Organic{}, \Enhanced{}, and \Suppressed{} realizations of the galaxy to the present day using the \Eagle{} model and adopt the `Recal' parameters that were shown by \citet{schaye_eagle_2015} to yield a better reproduction of the calibration diagnostics at the resolution of our zoom-in \ics{}. 

\Eagle{} uses stochastic implementations of several processes treated with subgrid models. When considering the ensemble properties of a large sample of galaxies, the random scatter introduced by stochastic implementations is effectively sampled over by the large number of objects, yielding outcomes influenced by both stochastic and causal mechanisms \citep{genel_quantification_2019}. However, this is not the case when studying individual galaxies on an object-by-object basis, where the variability introduced by the choice of random number seed can be significant \citep{keller_chaos_2019,davies_quenching_2021,borrow_impact_2023}. Therefore, we simulate the three sets of \ics{} nine times each, adopting a different seed to initialize the random number generator on each occasion. This enables a comparison of the scale of the effects stemming from the modified \ics{} to be compared with the scatter induced by the choice of the random number seed. Hereafter, we refer to the set of nine simulations corresponding to a given realization of the halo's assembly history as a `family'.

\begin{figure*}%
    \centering%
	\includegraphics[width=\textwidth]{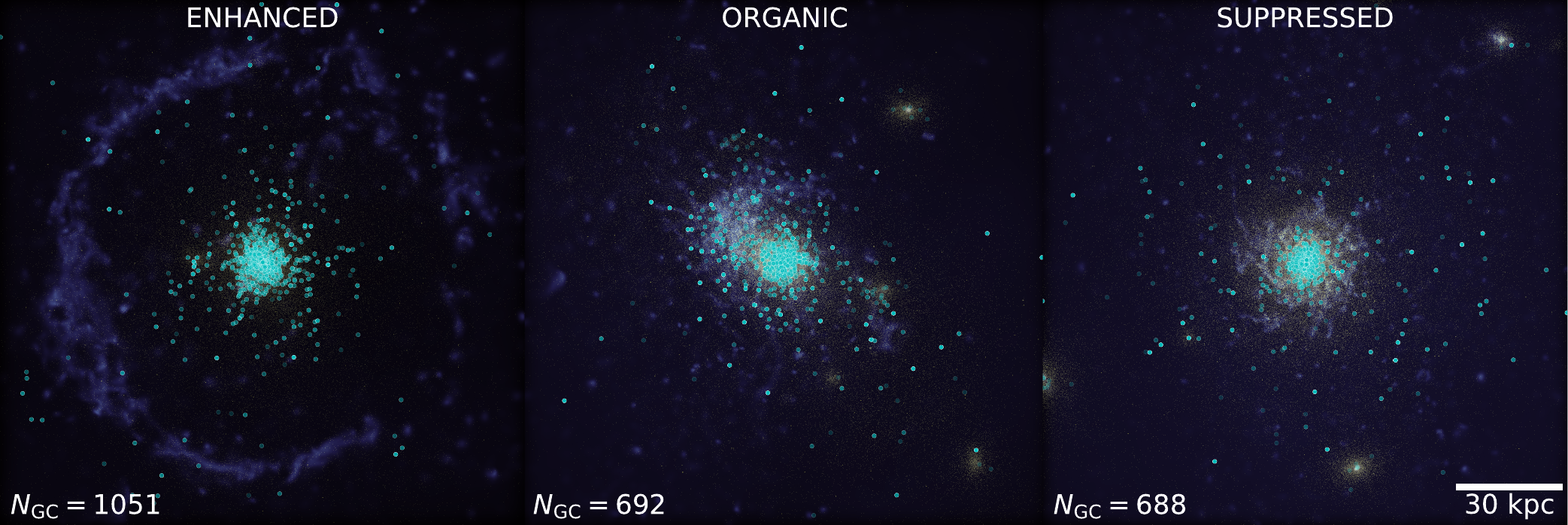}%
	\caption{The projected mass-weighted densities of gas~(blue), and stars~(yellow) within a \kpc[150\times150\times150] volume centred on the galaxy at \z[0] in the \Enhanced{}, \Organic{}, and \Suppressed{} cases (from left to right). The gas distribution is strongly influenced by feedback instigated by the major mergers, most notably in the \Enhanced{} case where a shell of ejected gas is visible. We mark the projected positions of \gc{}s~($\Mclust{}\geq\Msun[{10^5}]$) with circular cyan markers, and linearly decrease their alpha channel as a function of distance from the observer. Whereas over \percent{83} of the \gc{}s in the \Enhanced{} and \Suppressed{} cases are within \kpc[15] of the galaxy, in the \Organic{} case this figure is only \percent{67}. We indicate the total number of \gc{}s within \aperture{} of each galaxy at the bottom left of each panel.
	}%
	\label{fig:Results:galaxy_z0_render}%
	\vspace{-10pt}%
\end{figure*}%

\begin{table}%
	\centering%
 	\caption{Characteristics of the major mergers the galaxy experiences in each case (\Enhanced{}, \Organic{}, and \Suppressed{}). We show the median stellar mass ratio, $\mu_\ast\!\left(< \aperture{}\right),$ of the first and target mergers, and the \percent{68} scatter from the nine simulations comprising each family. We also present the lookback times by which the galaxies coalesce determined by visual inspection of the high-cadence simulation from each assembly history.}%
	\begin{tabularx}{\columnwidth}{YYYYY}%
	    \hline
	    & \multicolumn{2}{>{\hsize=\dimexpr2\hsize+2\tabcolsep+1\arrayrulewidth}Y}{First merger} & \multicolumn{2}{>{\hsize=\dimexpr2\hsize+2\tabcolsep+1\arrayrulewidth}Y}{Target merger} \\
	    \multirow{2}{*}{Simulation} & \multirow{2}{*}{$\mu_\ast$} & \tlb{} & \multirow{2}{*}{$\mu_\ast$} & \tlb{} \\
	    &  & $\left(\!\Gyr{}\right)$ &  & $\left(\!\Gyr{}\right)$\\
	    \hline
	    \Enhanced{}   & $0.65^{+0.02}_{-0.03}$ & $9.85$ & $0.36\pm0.06$ & $7.49$ \\
	    \Organic{}    & $0.45\pm0.03$ & $10.0$ & $0.16^{+0.04}_{-0.03}$ & $6.89$ \\
	    \Suppressed{} & $0.45\pm0.04$ & $9.95$ & --- & --- \\
	    \hline
	\end{tabularx}%
	\label{tab:Results:merger_information}%
	\vspace{-4pt}
\end{table}%

We output $29$~snapshots between \z[30] and 0, and additionally we run one simulation from each family with a higher output cadence such that it produces $55$~snapshots. We identify dark matter haloes within each snapshot by applying a friends-of-friends~(FOF) algorithm to the dark matter particle distribution, adopting a linking length of $0.2$~times the mean interparticle separation \citep{davis_evolution_1985}. Baryonic particles are associated to the FOF group, if any, of their nearest dark matter particle. Subsequently, we decompose haloes into gravitationally self-bound substructures using the \subfind{} algorithm \citep{springel_populating_2001,dolag_substructures_2009}. In general, we define the baryonic properties of the galaxy by aggregating the properties of particles located within \ckpc[30] of its centre of potential. We reconstruct the merger history of the galaxy by identifying its progenitor haloes in each preceding snapshot.

In \tabref{tab:Results:merger_information}, we list the median stellar mass ratios of the mergers and the \percent{68} scatter from each family of simulations. We supplement this information with the times by which the galaxies coalesce within the high-cadence simulation from each assembly history. The mass ratios of the first mergers in the \Organic{} and \Suppressed{} cases are similar; however, the first major merger in the \Enhanced{} case is more significant. Changes such as these, and other minor alterations to the overall accretion history of the halo, are a necessary component of the genetic modification approach in order to satisfy the constraint we impose that the halo mass at \z[0] must be similar in each assembly history. We discuss this change to the \ics{} in \secref{sec:Sum_Disc}.

\section{Results}
\label{sec:Results}
In \secref{sec:Results:Galaxy_properties}, we verify that several present-day properties of the galaxy in the three families of simulations are similar, and contrast the response of these properties to the modifications we make to the assembly histories. In \secref{sec:Results:GC_form_ev}, we examine the evolution of the stellar and \gc{} populations, including the disruption of the latter, of the galaxy in response to the modifications. In \secref{sec:Results:obs_measures_gcs} we characterize the richness of the \gc{} populations in each family of simulations using two popular observational metrics. In \secref{sec:Results:gc_z0_properties} we examine the fractional contributions of star clusters to the present-day stellar halo of the galaxy.

\subsection{Galaxy and halo properties}
\label{sec:Results:Galaxy_properties}
Our objective is to examine the influence of halo assembly history on \gc{} populations for haloes of fixed mass, so first we verify that our modifications to the \Organic{} \ics{} do not significantly change the present-day halo mass, \Mhalo{}. We impose no constraints on the baryonic properties of the galaxy because we wish to examine the influence of the dark matter assembly history on these quantities. We present \Mhalo{} and the stellar properties of the galaxies at \z[0] in \tabref{tab:Results:z0_properties}, which shows that the present-day halo mass of the galaxy in the three sets of simulations changes by only \percent{6} and is thus largely unaffected by the modifications of the \ics{}, as intended.

\begin{table}%
	\centering%
 	\caption{Present-day properties of the galaxy and its halo in each case: the halo mass, \Mhalo{}; the mass in stars, \Mstar{}; and the mass in \gc{}s, \Mgc{}, within \aperture{} of the galaxy. Median values and the \percent{68} scatter from the nine simulations (each adopting a different random number seed) corresponding to each assembly history are quoted.}%
	\begin{tabularx}{\columnwidth}{YYYY}%
	    \hline
	    \multirow{2}{*}{Simulation} & \Mhalo{} & $\Mstar{}\!\left(<30\, {\rm kpc}\right)$ & $\Mgc{}\!\left(<30\, {\rm kpc}\right)$ \\
	    & $\left(\Msun[10^{12}]\right)$ & $\left(\Msun[10^{10}]\right)$ & $\left(\Msun[10^{8}]\right)$\\
	    \hline
	    \Enhanced{}   & $3.1\pm0.1$ & $3.5^{+0.3}_{-0.5}$ & $6.7^{+1.1}_{-0.7}$ \\
	    \Organic{}    & $3.3\pm0.1$ & $4.5^{+0.2}_{-0.4}$ & $5.1^{+1.5}_{-0.9}$ \\
	    \Suppressed{} & $3.5\pm0.1$ & $3.8^{+0.3}_{-0.2}$ & $3.5^{+0.9}_{-0.5}$ \\
	    \hline
	\end{tabularx}%
	\label{tab:Results:z0_properties}%
	\vspace{-4pt}
\end{table}%

The stellar mass, \Mstar{}, and the mass in \gc{}s, \Mgc{}, respond naturally to the modifications made to the \ics{}. Consequently, the variation of these quantities with respect to the \Organic{} case is significant. The stellar mass of the galaxy in the \Enhanced{} and \Suppressed{} cases is \percent{{15-30}} lower than in the \Organic{} case, and \Mgc{} varies with respect to the \Organic{} case by \percent{{\pm30}}. The dichotomy in the behaviour of the unclustered stellar population and the evolution of the \gc{}s in response to the modification of the assembly history is unintuitive, so we discuss in detail the processes underpinning these outcomes in \secref{sec:Results:GC_form_ev}. The fractional scatter of the three properties listed in \tabref{tab:Results:z0_properties} increases as the columns are traversed from left to right, and is consistent with previous work indicating that stochastic variability in `cumulative' galaxy properties such as \Mstar{} scales with Poisson-like characteristics \citep{keller_chaos_2019,borrow_impact_2023}. The greater fractional scatter of \Mgc{} relative to \Mstar{} in each case stems from the smaller number of tracers.

In \figref{fig:Results:galaxy_z0_render}, we show the present-day projected mass-weighted densities of gas and stars in a representative realization from each family of simulations. We show the projected positions of the \gc{}s, i.e. clusters with $\Mclust{}\geq\Msun[{10^5}]$, using circular cyan markers. The modified assembly histories affect the structure of the galaxy. Particularly striking is the differing distribution of the gas: in the \Enhanced{} case most of the gas within \kpc[30] has been evacuated, unlike in the \Organic{} and \Suppressed{} cases which retain a reservoir of centrally concentrated cold gas. For this choice of random seed, the \Organic{} galaxy exhibits the most diffuse radial distribution of \gc{}s, while in the other two cases the \gc{}s are more concentrated. We find a similar result when comparing each family of simulations (not shown), although the difference is less severe. These results are consistent with those of other studies using particle tagging techniques to track the spatial evolution of the \gc{} population during galaxy mergers \citep[e.g.][]{renaud_origin_2017,chen_catalogue_2024}. The stellar morphology of the galaxy can be quantified in terms of the fraction of the total kinetic energy of the stars that contributes to corotational motion, \kcostar{} \citep[following][see also \citealt{thob_relationship_2019}]{correa_relation_2017}. We calculate \kcostar{} for all stellar particles within \aperture{} of the galaxy. In the \Suppressed{} case the galaxy exhibits a dominant disc component, with $\kcostar[0.50^{+0.04}_{-0.11}]$. In the other families of simulations the second (target) merger disrupts the ordered rotation, and \kcostar{} decreases to $0.33^{+0.07}_{-0.05}$ (\Organic{} case) and $0.24^{+0.02}_{-0.03}$ (\Enhanced{} case) as the significance of the second major merger increases.

\begin{figure}%
    \centering%
	\includegraphics[width=\columnwidth]{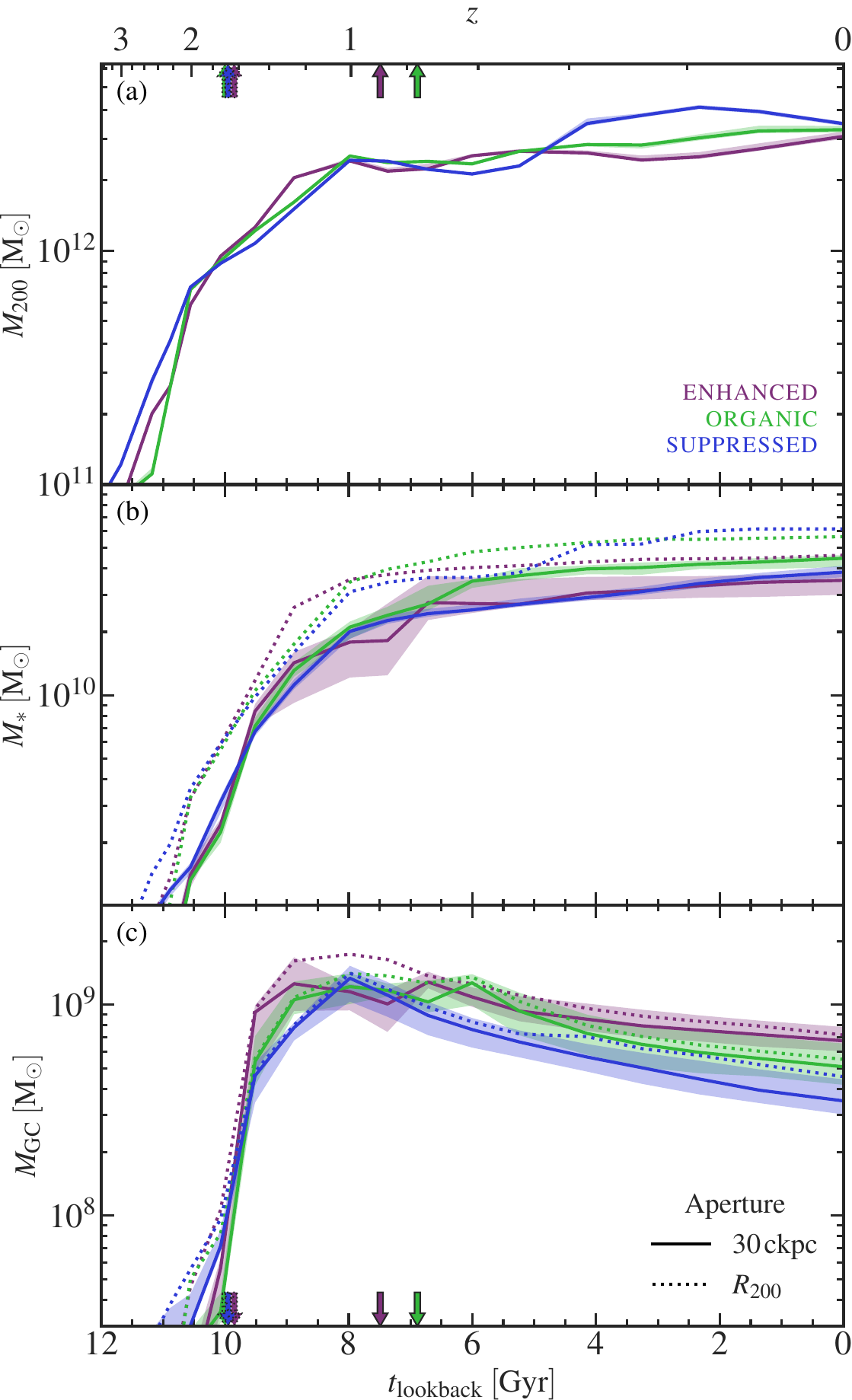}%
	\caption{The assembly histories of the galaxy for the three cases (\Enhanced{}, \Organic{} and \Suppressed{}) as a function of lookback time, \tlb{}, with the corresponding redshift shown on the upper horizontal axis. We show the evolution of the halo mass, \Mhalo{} (panel~a); and the mass of stars, \Mstar{} (panel~b), and \gc{}s, \Mgc{} (panel~c), within \aperture{} (solid curves) and \Rvir{} (dotted curves) of the galaxy's centre of potential. The curves show the median values of each quantity drawn from the nine simulations (using different random number seeds) comprising the family corresponding to each assembly history, and shaded regions show the corresponding \percent{68} scatter. Arrows with dotted outlines show the time of coalescence of the first significant merger experienced by the galaxies. The arrows with solid outlines show the time at which the galaxies completed the second significant merger, the mass ratio of which we alter with the genetic modification technique. The arrow corresponding to the target merger in the \Suppressed{} case is absent because no merger takes place by \z[0].
	}%
	\label{fig:Results:galaxy_properties}%
	\phantomlabel{panel~a}{fig:Results:galaxy_properties:M200}%
	\phantomlabel{panel~b}{fig:Results:galaxy_properties:Mstar}%
	\phantomlabel{panel~c}{fig:Results:galaxy_properties:Mgc}%
	\vspace{-10pt}%
\end{figure}

The top, centre and bottom panels of \figref{fig:Results:galaxy_properties} show, respectively, the temporal evolution of \Mhalo{}, $\Mstar{}\!\left(r<\aperture{}\right)$ and $\Mgc{}\!\left(r<\aperture{}\right)$, with solid curves denoting the median values of each quantity from the nine simulations comprising each family of \ics{}. Shaded regions show the corresponding \percent{68} scatter.\footnote{If a simulation from a family presents an invalid value, for example if there are no \gc{}s present in a particular snapshot and the quantity of interest cannot be computed, we exclude that snapshot when calculating the 16th, 50th, and 84th percentiles. If the percentile lies between values in the ranked list, we take the geometric mean of the bracketing values.} Where applicable, dotted curves denote the median value of the same quantity measured within \Rvir{} of the galaxy's centre of potential rather than within \aperture{}, and we adopt this convention throughout unless stated otherwise in the text. The modified assembly histories leave the early evolution of \Mhalo{} largely unaffected, which is as expected because the modifications do not explicitly target the first significant merger (at \z[1.7]). However, the modifications unavoidably and subtly affect other aspects of the halo assembly history. One example of this occurs in the \Enhanced{} case at \z[1.4], shortly after the first merger, when the evolution of \Mhalo{} briefly deviates from the \Organic{} case. Thereafter, despite the increased significance of the target merger in the \Enhanced{} case, \Mhalo{} tracks the mass assembly history of the \Organic{} case closely. In contrast, the absence of the final target merger in the \Suppressed{} case has a more significant effect on the evolution of \Mhalo{}, which deviates significantly from the \Organic{} halo assembly history after \tlb[{\Gyr[5]}]~$\left(z\simeq 0.5\right)$. Over the subsequent \Gyr[3], \Mhalo{} increases by nearly a factor of two, peaking at approximately \percent{30} above the present-day value of \Mhalo{}. This is caused by the flyby of a second, similarly massive halo that fails to merge into the halo of the target galaxy. The impact parameter of the interaction is sufficiently large that $\Mstar{}\!\left(r{<}\aperture{}\right)$ and $\Mgc{}\!\left(r{<}\aperture{}\right)$ are unaffected.

\begin{figure}%
    \centering%
	\includegraphics[width=\columnwidth]{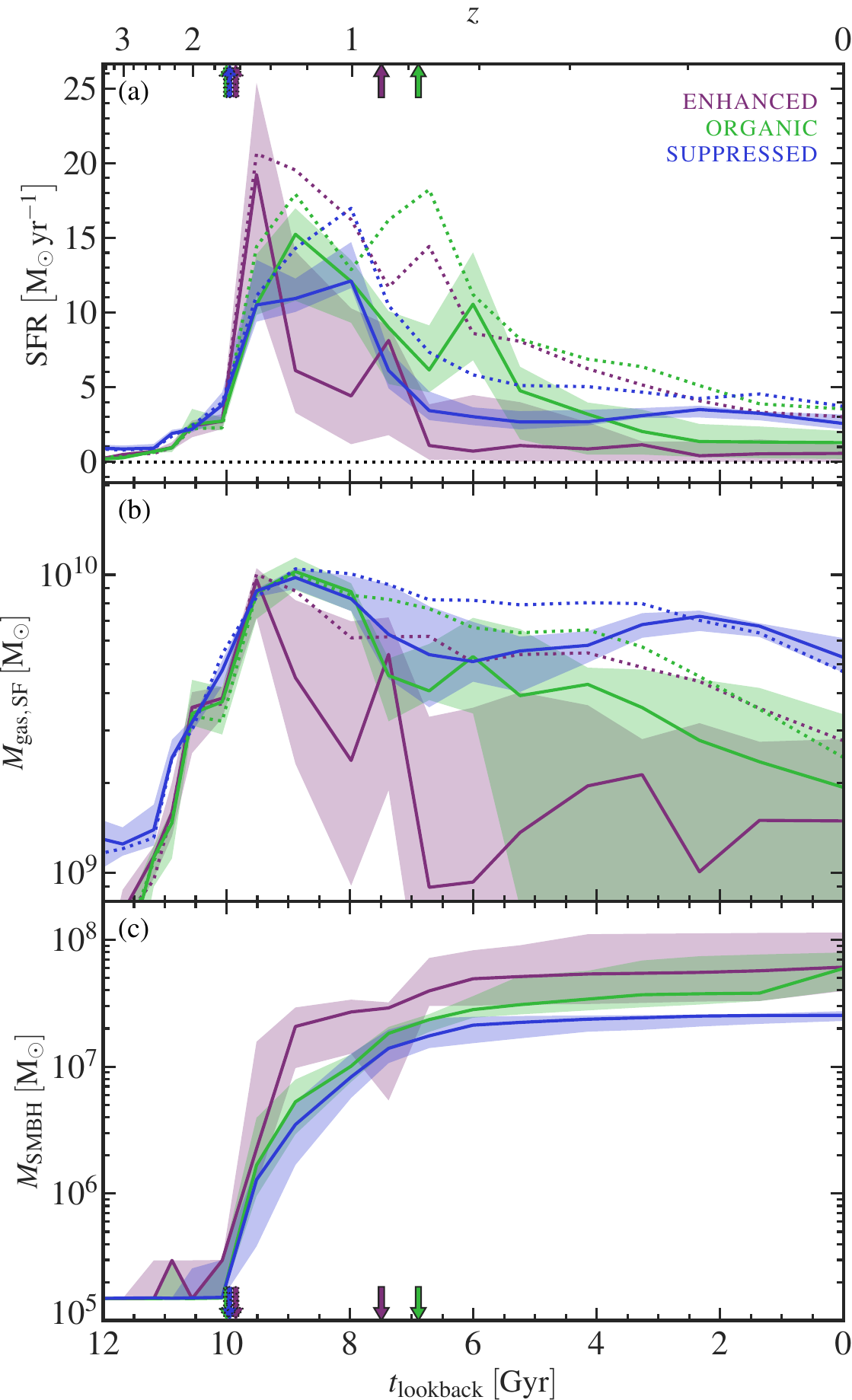}%
	\caption{The evolution of the star formation rate (panel a) and the mass in star-forming gas, \MgasSF{} (panel b), within \aperture{} of the centre of the galaxy. Solid curves correspond to the fiducial simulations, and dotted curves correspond to matched simulations in which AGN feedback is disabled. In panel~(c) we show the mass of the \SMBH{}, \Msmbh{}. As in previous figures, the hatched and solid arrows indicate the first and final significant mergers of each assembly history, respectively.
	}%
	\label{fig:Results:galaxy_mgassf_mbh}%
	\phantomlabel{panel~a}{fig:Results:galaxy_sfr_gfr:SFR}%
	\phantomlabel{panel~b}{fig:Results:galaxy_mgassf_mbh:MgasSF}%
	\phantomlabel{panel~c}{fig:Results:galaxy_mgassf_mbh:MSMBH}%
	\vspace{-10pt}%
\end{figure}

The evolution of the mass of stars (\figref{fig:Results:galaxy_properties:Mstar}) exhibits a more significant and enduring response to the modified assembly histories. Prior to \tlb[{\Gyr[8]}]~($z \simeq 1$), the stellar mass of the galaxy follows a common evolutionary path in all three cases. After this time, which is approximately coincident with the final significant merger in the \Enhanced{} case, both modified cases exhibit lower star formation rates than the \Organic{} case. The mechanisms underpinning this common feature differ between the two modified assembly histories. As we show in subsequent figures, in the \Enhanced{} case the conversion of baryons into stars is less efficient than in the \Organic{} case, owing to the more efficient growth of the \SMBH{} and the associated injection of additional energy from the active galactic nucleus~(AGN), which drives outflows capable of regulating the galaxy's fuel supply. In the \Suppressed{} case, episodes of strong AGN feedback are absent because the \SMBH{} grows less efficiently. In this version of the assembly history the star formation rate is reduced due to the lower characteristic pressure of the interstellar gas, which itself follows from a reduced gas infall rate. We examine both mechanisms that suppress the growth of stellar mass in more detail in \figref{fig:Results:galaxy_mgassf_mbh}.

In contrast to the stellar mass, the total mass of \gc{}s (\figref{fig:Results:galaxy_properties:Mgc}) yields a strong, monotonic response to the assembly history modifications. This emerges shortly after the first significant merger and prior to the peak in \Mgc{}. In all cases, \Mgc{} is enhanced by each significant merger event and the overall peak is reached approximately \Gyr[0.5] after the final significant merger the galaxy experiences. In the \Suppressed{} case the first merger is the only significant merger that the galaxy undergoes, so the peak of $\Mgc{}=\Msun[{1.34\times10^9}]$ is higher and is reached earlier (at \tlb[{\Gyr[8]}]) than for the \Organic{} and \Enhanced{} cases. The late-occurring final merger in those families fosters \percent{5} lower peak values of \Mgc{} at \tlb[{\Gyr[6]}]~($z\simeq 0.6$) and \tlb[{\Gyr[6.7]}]~($z\simeq 0.7$), respectively. In all cases the peak value of $\Mgc{}\!\left(t\right)$ is reached quickly at a relatively early epoch, and is followed by a prolonged decline of the mass of the \gc{} population that is sustained to the present day. Primarily, this is driven by the dynamical disruption of clusters, which moves stellar mass to the unclustered component of the stellar particles. We discuss the evolution of the \gc{} population and the processes that underpin it in more detail in \secref{sec:Results:GC_form_ev}. We note that while the ratio of \Mgc{} to \Mhalo{} is a factor of several higher than that found by \citet{harris_galactic_2017}, it is within the $68$ percentile scatter found in the \emosaics{} simulations \citep{bastian_globular_2020}.

To obtain a clearer view of the processes governing the growth of the target galaxy, we show in \figref{fig:Results:galaxy_mgassf_mbh} the evolution of the star formation rate of the galaxy, the mass of star-forming gas associated with the galaxy's main progenitor, \MgasSF{}, and the mass of the main progenitor's \SMBH{}, \Msmbh{}. The first significant merger in the \Enhanced{} case triggers an intense episode of star formation that peaks at almost \Msunyr[20], after which the rate drops precipitously. This coincides with a rapid decline in the availability of star-forming gas brought about by outflows driven by young stellar populations and AGN. The target merger contributes a small amount of cold gas and briefly elevates the star formation rate. Thereafter, the rate of star formation declines as the cold gas is consumed or expelled. In the \Suppressed{} case the star formation rate is also elevated by the first significant merger. The peak formation rate of \Msunyr[12] is much lower than the peaks in the \Organic{} and \Enhanced{} cases; however, it is sustained for approximately \Gyr[1.5] longer because the \SMBH{} grows more slowly and the feedback energy it injects is insufficient to regulate star formation. Thus, most of the gas that is accreted during the merger remains available for star formation throughout this time. From \tlb[{\Gyr[6.5]}] ($z \simeq 0.7$) to the present day, the galaxy in the \Suppressed{} case forms stars at a reasonably constant rate of approximately \Msunyr[3.5]. Over the same period the star formation rate in the \Enhanced{} case is approximately \Msunyr[0.8], which is a factor of more than four lower. This significant disparity in the behaviour of the star formation rates at late times enables the galaxy with the \Suppressed{} assembly history to form a greater mass of field stars by \z[0] than the galaxy in the \Enhanced{} case (see \figref{fig:Results:galaxy_properties:Mstar}).

\begin{figure}%
    \centering%
	\includegraphics[width=\columnwidth]{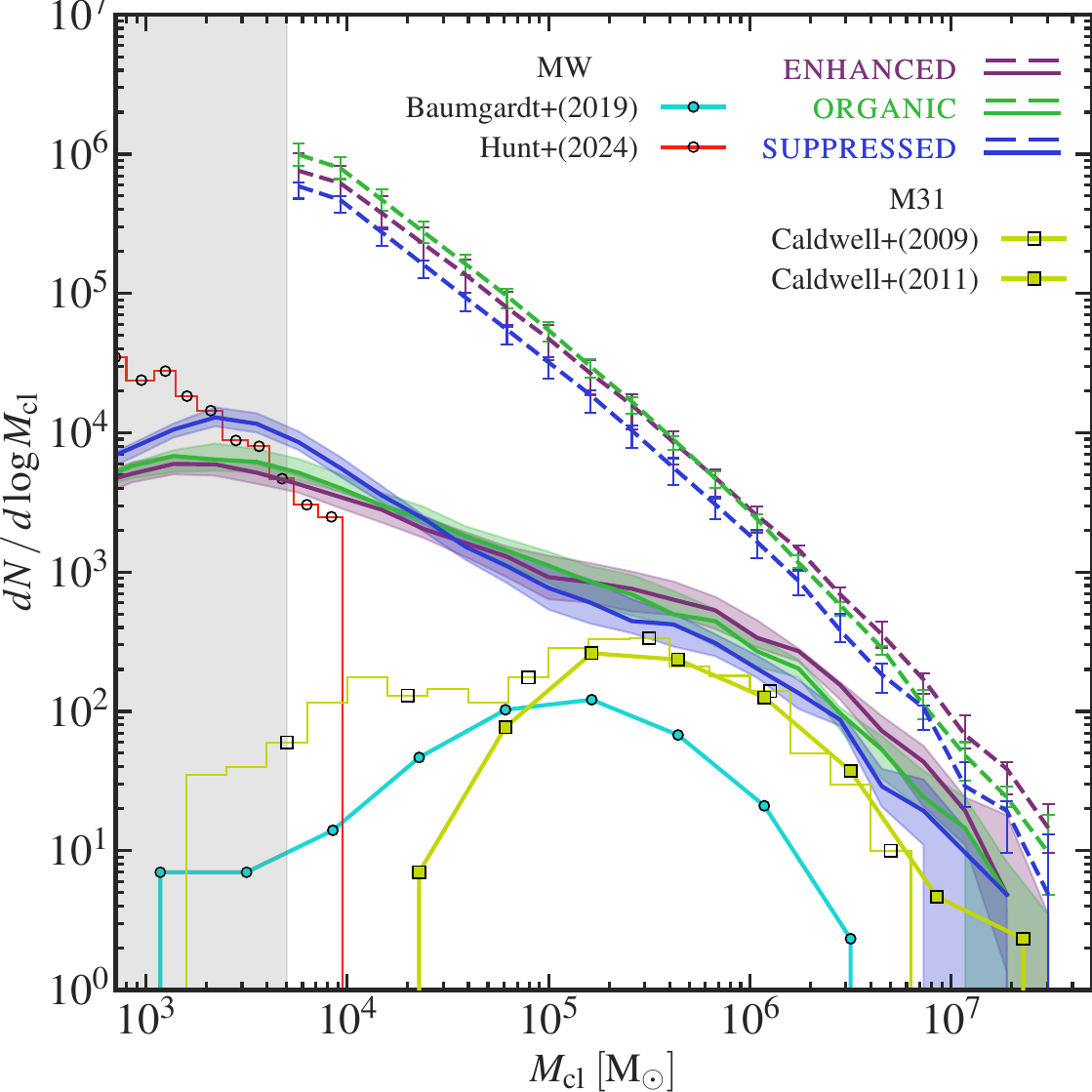}%
	\caption{The mass functions of all star clusters surviving at \z[0] in the three families of simulations. The solid curves and shaded regions show the medians and \percent{68} scatter of the \z[0] mass functions of the star clusters. Dashed curves and error bars show the equivalent quantities for the birth mass functions of all star clusters, whether surviving or fully disrupted, associated with star particles within the virial radius. The latter are a mass-weighted integral over all clusters that formed in the galaxy and are not equivalent to the cluster birth mass function used to seed the cluster population in the \emosaics{} subgrid models (see \secref{sec:Methods:Emosaics} for details). In our simulations, clusters that form with initial masses below \Mclust[{\Msun[{5\times10^3}]}], which we show with a shaded region at the left of the figure, are not tracked because they disrupt rapidly. For comparison we overlay the catalogues of Milky Way cluster data collated in \citet[][and subsequently updated]{baumgardt_mean_2019}, the completeness-corrected open cluster data from \citet{hunt_improving_2024}, and the M31 cluster data collated in \citet{caldwell_star_2009,caldwell_star_2011}.
	}%
	\label{fig:Results:galaxy_z0_cluster_mass_function}%
\end{figure}%

The availability of cold gas in the vicinity of the galaxy can be reduced by AGN feedback driven by merger events \citep[see e.g.][]{davies_galaxy_2022}. We demonstrate how this process influences our system in \figref{fig:Results:galaxy_mgassf_mbh}, panels~b and~c. In all three cases the mass of star-forming gas within \aperture{} of the centre of the galaxy's main progenitor peaks at approximately \Msun[{10^{10}}] shortly after the first significant merger. Thereafter, the star-forming gas reservoir in the \Enhanced{} case is rapidly reduced by AGN-driven outflows, and by a burst of star formation triggered by the merger (discussed further in \secref{sec:Results:GC_form_ev}). The AGN feedback effects are illustrated most clearly by comparing the evolution of the mass of star-forming gas in the fiducial simulations (solid curves) with that which emerges in counterpart simulations in which AGN feedback has been disabled (dotted curves): the mass of star-forming gas is suppressed by AGN feedback most strongly in the \Enhanced{} case. In this case the \SMBH{} is able to grow efficiently, reaching a mass of $\Msmbh{} \simeq \Msun[{2\times 10^7}]$ shortly after \tlb[{\Gyr[9]}] ($z \simeq 1.3$), powering energetic winds that expel most of the remaining gas over the subsequent \Gyr[4]. The final significant merger at \tlb[{\Gyr[7.5]}] ($z \simeq 0.9$) provides a fresh supply of cold gas that briefly interrupts the depletion of the gas reservoir; however, this gas is quickly consumed or expelled. The two mergers experienced by the \Organic{} galaxy are less significant, and as a result the SMBH grows more slowly and more star-forming gas is retained. Removing the target merger entirely in the \Suppressed{} case causes its SMBH to grow more slowly still, and to a lower final mass, yielding higher \MgasSF{} at \z[0].

In this section we have shown that the  evolution of \Mgc{} depends on the assembly history of the host galaxy. Galaxies experiencing two significant mergers have higher \Mgc{} at \z[0] than galaxies experiencing only one. The growth of \Mgc{} is correlated with star formation rates, which are enhanced during the mergers, and the availability of cold star-forming gas, which is depleted by a combination of star formation, and stellar and AGN feedback.

\subsection{Globular cluster formation and evolution}
\label{sec:Results:GC_form_ev}

\figref{fig:Results:galaxy_z0_cluster_mass_function} shows the present-day mass functions of surviving star clusters with mass $\Mclust{}\geq\Msun[10^2]$ within the galaxy's virial radius, \Rvir{} (solid curves), and the initial `birth' mass function of all clusters (dashed curves), whether surviving or fully disrupted, associated with stellar particles within \Rvir{}. For comparison, we also overlay the observed cluster mass functions of M31 \citep{caldwell_star_2009,caldwell_star_2011} and the Milky Way \citep{baumgardt_mean_2019}, and the recent completeness-corrected catalogue of open clusters in the Milky Way \citep{hunt_improving_2024}. Below \Mclust[{\Msun[10^4]}], the Milky Way open cluster data follow a similar trend to the low-mass component of the simulated cluster mass functions. The apparent cut-off in the former at \Mclust[{\Msun[10^4]}] is because the observations probe only a small volume around the Sun. The most massive known young star clusters in the Milky Way have masses $\Mclust \simeq \Msun[10^5]$ \citep{clark_massive_2005,figer_discovery_2006,davies_massive_2007,alexander_discovery_2009,davies_glimpse-co1_2011}. For $\Mclust{}\gtrsim\Msun[10^5]$ the observational data have a similar shape to the cluster mass functions exhibited by all three assembly histories of the simulated galaxy.

The cluster birth mass functions of each realization of the galaxy shown in \figref{fig:Results:galaxy_z0_cluster_mass_function} represent a mass-weighted average of the cluster birth mass functions of all stellar particles within \ckpc[30] of the galaxy centre at \z[0]. As we do not track the evolution of star clusters that form with initial masses at or below \Mclust[{\Msun[5\, 000]}], the birth mass functions shown in \figref{fig:Results:galaxy_z0_cluster_mass_function} are not populated below this threshold. We indicate the latter with a shaded region at the left of the figure. As discussed in \secref{sec:Methods:Emosaics}, the form of each stellar particle's cluster birth mass function depends on its natal gas conditions, so the galaxy averaged cluster birth mass functions are influenced by the evolving gas conditions of their progenitors. It is clear from the figure that the modifications to the galaxy's assembly history affect the cluster birth mass function: while the dashed curves have broadly similar shapes, the slope exhibited by the \Enhanced{} case is notably shallower, owing to the elevation of \Mcstar{} during the epoch of peak cluster formation. As a result, the \Enhanced{} assembly history promotes the formation of more high-mass clusters ($\Mclust{} > \Msun[10^5]$) and fewer low-mass clusters. It is also clear from the normalization of the curves that the total stellar mass formed in clusters is greater in the \Enhanced{} case than the \Organic{} case, which in turn is greater than the \Suppressed{} case, although the form of the cluster birth mass function is similar for the \Organic{} and \Suppressed{} cases.

\begin{figure}%
    \centering%
	\includegraphics[width=\columnwidth]{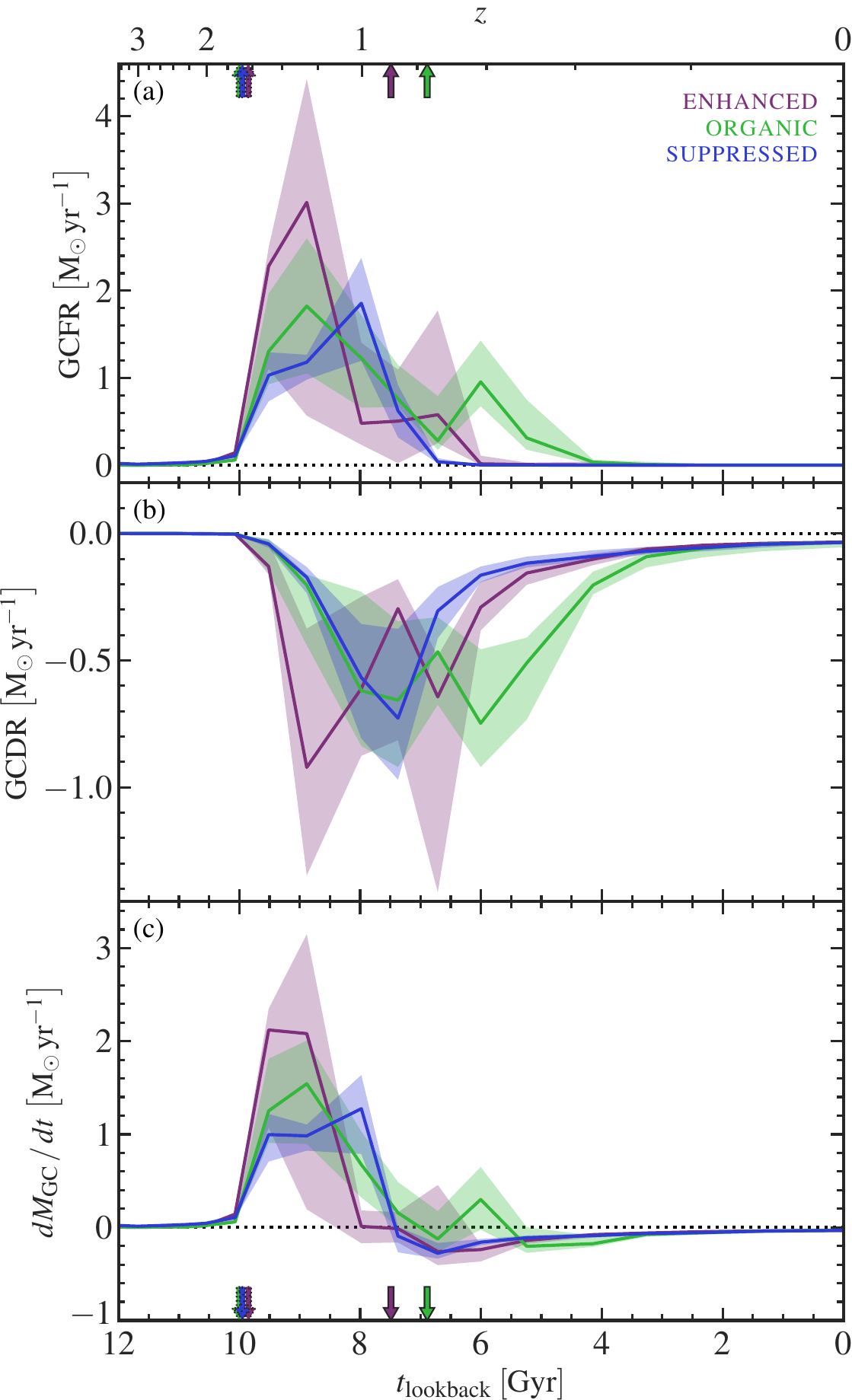}%
	\caption{The evolution of the \gc{} formation rate (panel a), the \gc{} disruption rate (panel b), and the net change in the mass of \gc{}s (panel c) within \aperture{} of the centre of the galaxy. The hatched arrows with dotted outlines show the time of the first significant merger experienced by the galaxies, and the arrows with solid outlines show the time at which the galaxies experience the second significant merger.}%
	\label{fig:Results:galaxy_sfr_gfr}%
	\phantomlabel{panel~a}{fig:Results:galaxy_sfr_gfr:GCFR}%
	\alphantomlabel{panel~a}{fig:Results:galaxy_sfr_gfr:GCFRal}%
	\phantomlabel{panel~b}{fig:Results:galaxy_sfr_gfr:GCDR}%
	\phantomlabel{panel~c}{fig:Results:galaxy_sfr_gfr:netdMgc}%
\end{figure}%

The surviving cluster mass function also exhibits a greater number of massive clusters in the \Enhanced{} case, indicating that the disruption or ejection of massive clusters stemming from this particular assembly history is insufficient to offset their formation. These modest differences between assembly histories are consistent with the results from the \emosaics{} simulations in which \percent{10-15} of surviving \gc{}s formed directly during major mergers \citep{keller_where_2020}. We examine this issue in more detail below. Relative to the other assembly histories, the surviving cluster mass function of the \Suppressed{} case exhibits a conspicuous excess of clusters with masses below \Mclust[{\Msun[2\times10^4]}]. This arises because the tidal field in the \Suppressed{} case changes more slowly than for the other assembly histories, and the absence of an explicit treatment of the multiphase interstellar medium in \Eagle{} causes the under-disruption of clusters. Consequently, some of the surviving low-mass cluster population at \z[0] originates from old, `over-surviving' clusters, and the remainder comprises young clusters that were born with low masses. The latter
experience less disruption than their counterparts in the \Organic{} and \Enhanced{} cases where tidal shocks induced by the target merger efficiently disrupt clusters. These effects combine to produce an over-abundance of low-mass clusters at \z[0] in the \Suppressed{} case. We discuss this in more detail in \secref{sec:Sum_Disc}.

\begin{figure}%
    \centering%
    \includegraphics[width=\columnwidth]{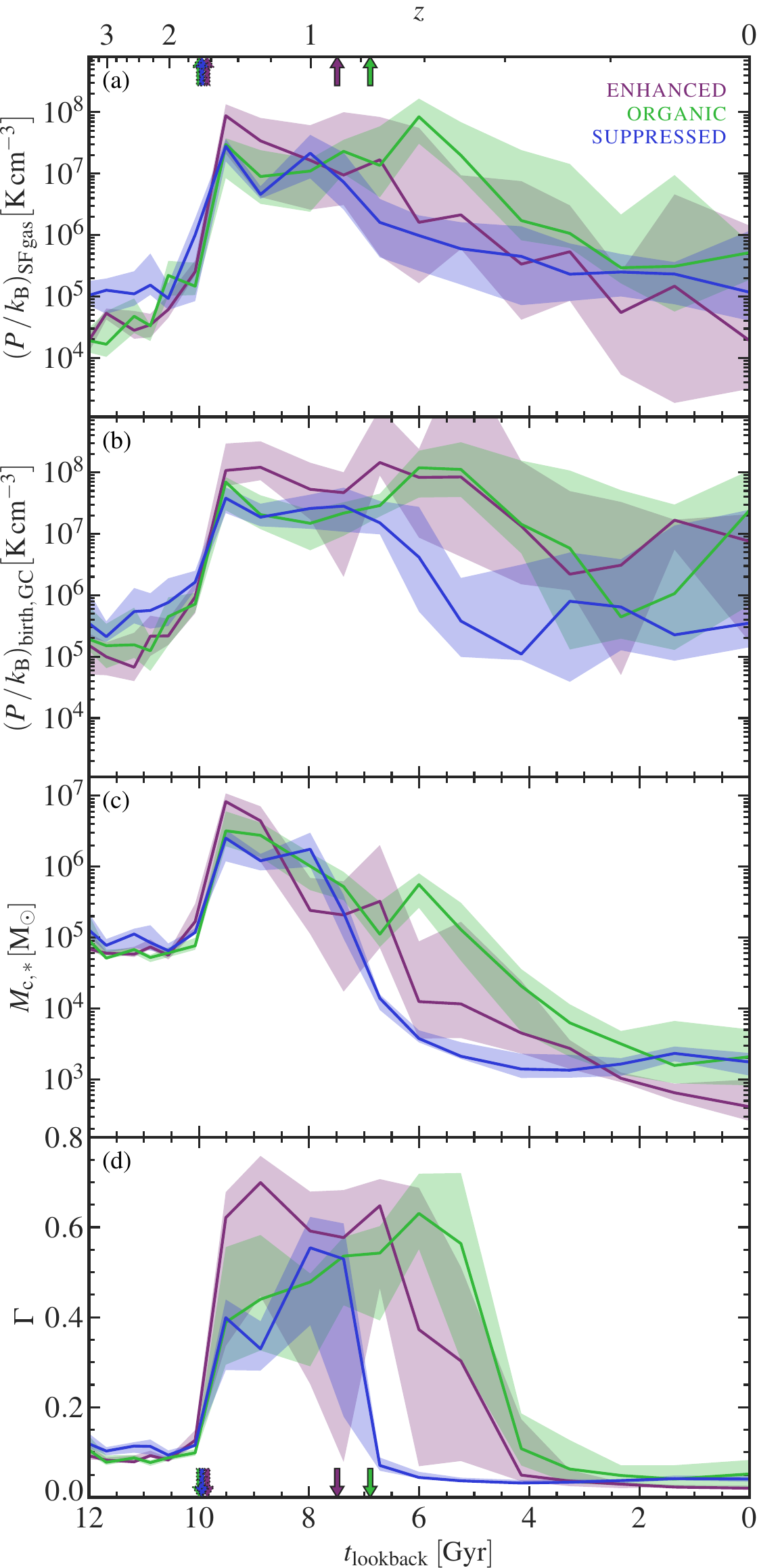}%
    \caption{The evolution of the star formation rate-weighted pressure of the star-forming gas, $\left(P\, /\, k_{\rm B}\right)_{\rm SF\, gas}$ (panel a); the stellar mass-weighted birth pressure of \gc{}s, $\left(P\, /\, k_{\rm B}\right)_{\rm birth,\, GC}$ (panel b); the median truncation mass of the initial cluster mass function, \Mcstar{} (panel c); and the median cluster formation efficiency, \CFE{} (panel d); within \aperture{} of the centre of the galaxy. These properties, which are determined by the properties of the natal gas, influence the creation of \gc{}s in the \mosaics{} model. As in previous figures, the arrows show the times of the first (hatched fills with dotted outlines) and final (solid outlines) significant mergers experienced by the galaxies.
    }%
	\label{fig:Results:mosaics_model_properties}%
	\phantomlabel{panel~a}{fig:Results:mosaics_model_properties:PkbirthStar}%
	\phantomlabel{panel~b}{fig:Results:mosaics_model_properties:PkbirthGC}%
	\phantomlabel{panel~c}{fig:Results:mosaics_model_properties:Mcstar}%
	\phantomlabel{panel~d}{fig:Results:mosaics_model_properties:CFE}%
	\vspace{-10pt}%
\end{figure}%

The formation rates of the \gc{}s, shown in \figref{fig:Results:galaxy_sfr_gfr:GCFR}, follow similar patterns to the star formation histories discussed in \secref{sec:Results:Galaxy_properties}. Although not shown in the figures, in all three assembly histories the oldest surviving \gc{}s formed as early as \tlb[{\Gyr[{13.4-13.5}]}]~(\z[{10-12}]). This is consistent with \citet{pfeffer_comparing_2025}, who showed that \gc{}s form as early as \z[10] in \emosaics{} galaxies; and with several other \gc{} formation and evolution models \citep[see][for a recent comparison]{valenzuela_globular_2025}. In the realizations of the \Lstar{} galaxy the \gc{} formation rate at this early time is approximately \Msunyr[{10^{-5}}], and it remains low until the first significant merger. This causes the most rapid increase in the rate of \gc{} formation, and the rate of change in the \gc{} formation rate scales with the mass ratio of the galaxies participating in the first merger. In the \Organic{} and \Enhanced{} cases, which experience a second significant merger, the coalescence of the merging galaxies also triggers a period with an elevated \gc{} formation rate. Following the mergers \gc{} formation shuts down completely, at lookback times of \Gyr[{4-6}]. This contrasts sharply with the formation of new field stars, which continues to the present day. The disparity in the formation rates of field stars and \gc{}s emerges because, in the \emosaics{} model, \gc{} formation requires high gas pressure. Such pressures are commonplace at early times; however, they are only produced at low redshift during significant merger events \citep{reina-campos_formation_2019}.

\figref{fig:Results:galaxy_sfr_gfr} shows that all three assembly histories share a common phase, starting after the first significant merger and ending at \tlb[{\Gyr[8]}]~($z\simeq 1$), during which \gc{}s form at a faster rate than they are disrupted. We define \gc{} `disruption' as the mass loss due to evaporation and shocks from clusters that formed as \gc{}s, i.e. with masses greater than \Mclust[{\Msun[{10^5}]}]. As panels~(a)~and~(b) show, \gc{} disruption is largely co-temporal with the elevated formation rates because most disruption takes place in the natal environment \citep{kruijssen_dynamical_2012}. In the \Organic{} and \Enhanced{} cases, which experience a second significant merger, the phase of net \gc{} formation is prolonged by \Gyr[{0.5-1}]. Following the cessation of \gc{} formation, there is a net reduction in the mass of the \gc{} population. During this phase, the \gc{} disruption rate decays gradually, and in all three assembly histories it settles at a low but non-negligible rate of approximately \Msunyr[0.1] over the final \Gyr[4] of galaxy evolution. The transition from the growth phase to the decay phase is governed by the availability and distribution of high-pressure star-forming gas in the centre of the galaxy, which differs significantly between the three assembly histories after \tlb[{\Gyr[8]}]. 

The transition also depends on several properties of the gas that are considered by the star cluster formation model discussed in \secref{sec:Methods:Emosaics} \citep[e.g.][]{kruijssen_globular_2015,kruijssen_formation_2019,reina-campos_formation_2019}. As we discussed in that section, the characteristic mass scale of the cluster birth mass function, $\Mcstar{}$, and the fraction of clustered star formation, \CFE{}, are both dependent on natal gas pressure. We therefore show in \figref{fig:Results:mosaics_model_properties} the star formation rate-weighted pressure of star-forming gas (panel a), and the stellar mass-weighted birth pressure of \gc{}s (panel b), along with the mass-weighted values of \Mcstar{} (panel c) and \CFE{} (panel d). The differences between the curves in panels~(a) and (b) reflect that \gc{} formation is biased towards the high-pressure regions of the interstellar medium. The first significant merger fosters a sharp increase, by several orders of magnitude, of the characteristic pressure of star-forming gas in all three assembly histories. Compared with the other assembly histories, the pressure of the interstellar medium increases most strongly in the \Enhanced{} case to counterbalance the deeper gravitational potential created by the higher mass-ratio merger. This boosts the average star-forming gas pressure to nearly $P/k_{\rm B} = $\KcmCubed[{10^8}], which is $10^3 - 10^4$ times higher than the average pressure in the interstellar medium before the merger. Such high gas pressures strongly favour the production of clustered star formation, particularly massive clusters, and this fosters the dramatic increase in the rate of \gc{} formation in the \Enhanced{} case shortly after the first significant merger (see \figref{fig:Results:galaxy_sfr_gfr:GCFRal}). In the \Suppressed{} case the star-forming gas pressure peaks at a similar time but at a lower value of \KcmCubed[{3\times10^7}]. After a \Gyr[2] period during which the average gas pressure remains elevated, it declines until the present day. In this scenario the first merger plays the most important role in governing the subsequent evolution of the gas. The shallower gravitational potential after the merger relaxes more quickly than in the other cases, dispersing cold gas more widely and reducing the star formation rate. In the \Organic{} case the galaxy reaches its peak in the average gas pressure at \tlb[{\Gyr[6]}] ($z \simeq 0.6$), which is much later than the other cases. This is caused by the final significant merger that provides the inner regions of the galaxy with a renewed supply of cold gas.

\begin{figure*}%
    \centering%
	\includegraphics[width=\textwidth]{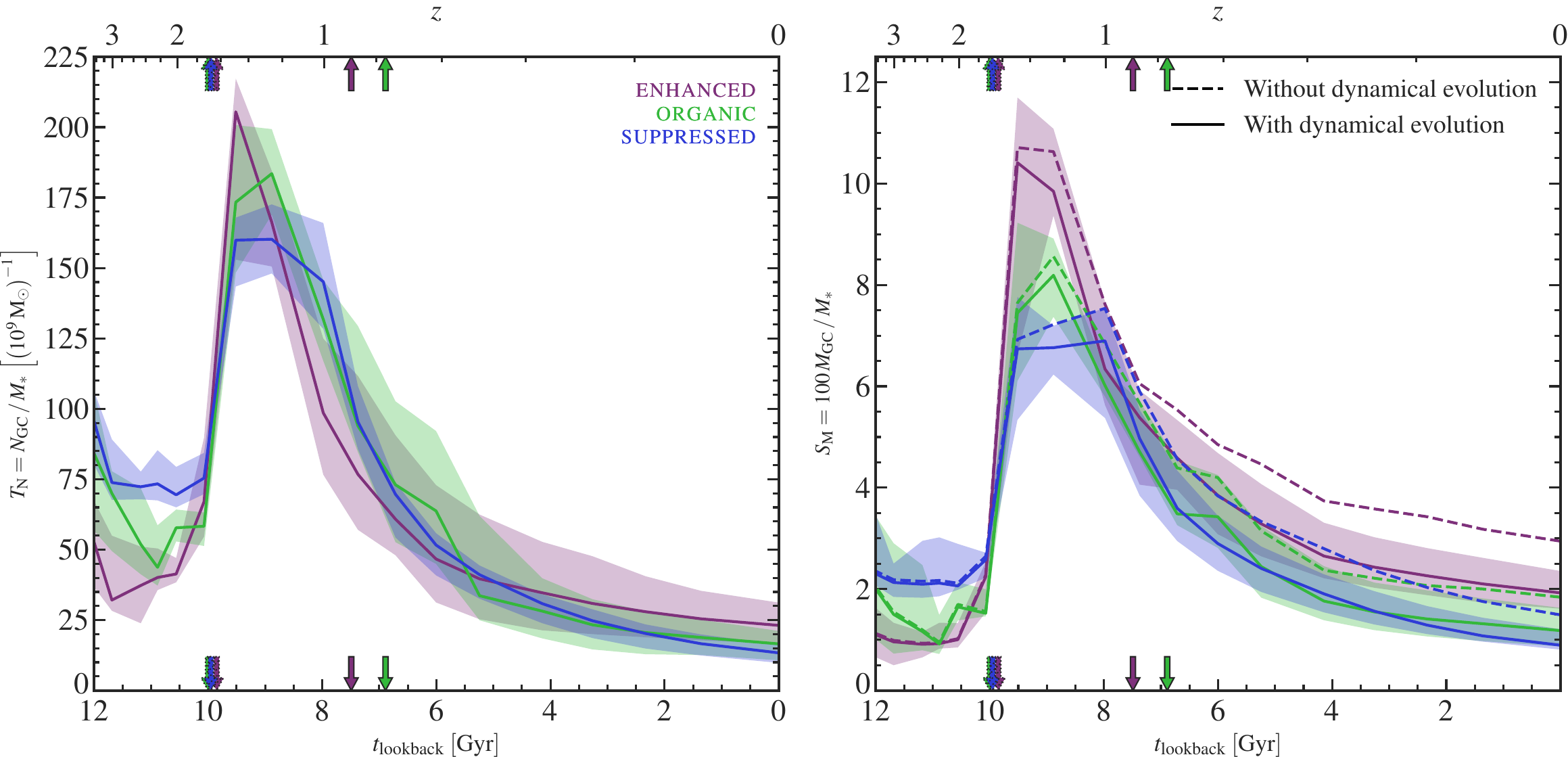}%
	\caption{The evolution of the specific frequency, $\TN{}=\Ngc{}\, /\, \Mstar{}$, and the specific mass, $\SM{} = 100\Mgc{}\, /\, \Mstar{}$ (left and right panels, respectively) within \aperture{} of the centre of the galaxy. These diagnostic quantities encode information about the past conditions that were favourable towards the production, and disruption, of \gc{}s compared with the overall star formation history of a galaxy. In the right panel, the solid curves are computed using the fiducial \mosaics{} cluster formation model, and the dashed curves show the evolution of \SM{} after removing any dependence on the dynamical evolution of the clusters to demonstrate the competing effects arising from the changes to the merger history.
	}%
	\label{fig:Results:galaxy_TN_SM}%
	\vspace{-10pt}%
\end{figure*}%

Panel~(c) of \figref{fig:Results:mosaics_model_properties} shows the evolution of \Mcstar{}, the truncation mass of the initial cluster mass function. The rapid pressurization of gas during the first significant merger increases the maximum mass of star clusters that can form by almost two orders of magnitude, peaking at \Msun[{10^6-10^7}] in all three assembly histories. This enables the formation of massive star clusters and stimulates the increase in the rate of \gc{} formation highlighted by \figref{fig:Results:galaxy_sfr_gfr}. The completion of the merger marks the start of a sustained decline in the value of \Mcstar{} that accelerates after \tlb[{\Gyr[8]}]~($z \simeq 1$) and continues until the present day. During this epoch the average truncation mass of the cluster birth mass function is \Mcstar[{\Msun[{10^3-10^4}]}], and the formation of \gc{}s with $\Mclust{}\geq{}\Msun[10^5]$ is strongly disfavoured. Consequently, \gc{} formation at these late times becomes highly stochastic.

The cluster formation efficiency, \CFE{}, shown in panel~(d) is closely related to the natal gas pressure and governs, on a per gas particle basis, the fraction of stars born in clusters. In the \Gyr[2] prior to the first significant merger, \CFE{} is relatively stable at a value of $\CFE{} \simeq 0.1$ in all three cases. The first merger rapidly pressurizes the gas in the inner \aperture{} to a peak of $P/k_{\rm B} >$\KcmCubed[{10^{7.5}}], increases the efficiency of cluster formation by factors of $3-7$, and triggers similarly rapid growth of the \SMBH{} (see the discussion of \figref{fig:Results:galaxy_mgassf_mbh} in \secref{sec:Results:Galaxy_properties}). The growth of the \SMBH{} is particularly abrupt in the \Enhanced{} case, and the associated feedback ejects a large amount of gas from the centre of the galaxy. The final merger sustains the high gas pressure in the \Enhanced{} case for a further \Gyr[1.5] before it drops below the pressure that prevailed at early times. The cluster formation efficiency tracks the evolution of the gas pressure closely, and \percent{{60-70}} of the stars that are born in this epoch are formed in star clusters.

In the \Organic{} and \Suppressed{} cases, the response to the first significant merger is more muted than in the \Enhanced{} assembly history. The pressurization of the star-forming gas takes place slightly more gradually, and \CFE{} increases slowly over the subsequent \Gyr[3-5]. In contrast to the \Enhanced{} case, the final merger in the \Organic{} case causes the already high gas pressures to increase by another order of magnitude, and extends the period of high-efficiency cluster formation by a further \Gyr[1]. The galaxy in the \Suppressed{} case lacks a second merger to sustain highly efficient cluster formation. Consequently, new cluster formation ceases almost completely shortly after \tlb[{\Gyr[7]}]~($z \simeq 0.8$), which is approximately \Gyr[3] after the merger.

In this section, we have shown that galaxies experiencing two significant mergers have more massive and fewer low-mass star clusters at \z[0] compared to galaxies in which the second major merger does not take place. The rates of \gc{} formation and disruption are both enhanced during major mergers, although \gc{}s form more quickly than they are disrupted until \Gyr[{0.5-1}] after the final major merger. This behaviour arises because the star-forming gas becomes pressurized during the merger, which increases \CFE{} and \Mcstar{} and boosts the formation of \gc{}s more strongly than it boosts their disruption.

\subsection{Characterizing the globular cluster system in terms of observational quantities}
\label{sec:Results:obs_measures_gcs}
The \gc{} richness of galaxies can be characterized in terms of the number, \Ngc{}, and mass of \gc{}s relative to the mass of stars. These quantities are known respectively as the specific frequency, $\TN{}=\Ngc{}\, /\, \Mstar{}$, and the specific mass, ${\SM{}=100\Mgc{}\, /\, \Mstar{}}$, and they encode information about the past conditions that favoured the formation of \gc{}s and trace their subsequent disruption relative to star formation in the galaxy and its progenitors. This makes them useful diagnostics with which to compare galaxies of different morphological types and, of particular interest for this study, with different assembly histories. The specific frequency is weighted towards lower mass \gc{}s, and correlates strongly with the burst of \gc{} formation that takes place during the first merger. Its subsequent evolution is governed primarily by disruption processes that deplete the \gc{} system in combination with low cluster formation efficiency at late times. By \z[0], \TN{} effectively traces the disruption of \gc{}s during the evolution of the galaxy. This result is consistent with that of \citet[][see their \extfig{4}, in particular]{kruijssen_globular_2015}. The specific mass is weighted towards more massive clusters and is therefore sensitive to both \CFE{} and \Mcstar{} \citep{kruijssen_globular_2015}. Larger values reflect favourable conditions for the formation of massive clusters. Of course, both quantities also encode a dependence on the rate of \gc{} disruption, which affects their subsequent evolution. As noted by \citet{peng_acs_2008} and \citet{bastian_globular_2020}, an advantage of \SM{} is that it is less susceptible to stochasticity than \TN{} in galaxies with few \gc{}s. 

The peak in \TN{} coincides with the completion of the first significant merger (see \figref{fig:Results:galaxy_TN_SM}, left panel). As surmised above, the high pressures that ensue in the star-forming gas following the first merger contribute to a significant increase in the efficiency of cluster formation in general, and \gc{} formation in particular. In all three assembly histories \TN{} remains elevated for approximately \Gyr[2] before returning, over a similar period of time, to the levels prevailing in the main progenitor prior to the mergers. This behaviour is controlled by the rate of \gc{} disruption in the galaxy, which dominates earlier in the \Enhanced{} assembly history than in the other cases. Similarly, after the first merger \Mcstar{} peaks at a higher value in the \Enhanced{} case than in the other two cases (see \figref{fig:Results:mosaics_model_properties:Mcstar}). This can be interpreted as fostering the formation of a \gc{} population with a higher average cluster mass. Consequently, the evolution of \SM{} becomes stratified shortly after the first merger and remains so until \z[0]. This behaviour is replicated more weakly in \TN{} because, unlike \SM{}, it is dominated by low-mass \gc{}s that rapidly evolve below our \gc{} mass threshold in response to the local environments in each assembly history. By \z[0], the values of \TN{} are similar and the different assembly histories are indistinguishable using this metric alone. In each case, at \z[0] the simulated galaxies exhibit values of \TN{} and \SM{} that are much larger than the corresponding present-day metrics calculated for the Milky Way and M31. These calculations are based on recent catalogues of star clusters \citep{caldwell_star_2011,baumgardt_mean_2019} and estimates of the stellar masses of these systems \citep{licquia_improved_2015,sick_stellar_2015}.

\figref{fig:Results:galaxy_TN_SM} shows that, for the galaxy assembly histories we consider, any effect the final merger may have on \TN{} or \SM{} is comparable to the stochasticity introduced by the choice of random number seed. Consequently, the intrinsic efficiency of \gc{} formation and disruption per unit baryonic mass (in stars and gas) is similarly unaffected. This behaviour arises primarily because the final merger contributes little of the integrated \gc{} production, nor does it enhance the disruption of \gc{}s significantly compared to the formation and disruption that happened at earlier times (see e.g. \figref{fig:Results:galaxy_sfr_gfr:netdMgc}). In a galaxy experiencing a less significant major merger at early times, \TN{} and \SM{} may exhibit stronger sensitivity to the late-time merger that could be significant compared to the stochasticity induced by the random number seed. Follow-up studies considering a variety of assembly histories would shed light on this.

Similarly, the second merger has less influence on the disruption of the galaxy's \gc{} population. The dashed curves in the right panel of \figref{fig:Results:galaxy_TN_SM} denote \SM{} calculated in the absence of dynamical mass loss, such that clusters only lose mass via stellar evolution. The solid and dashed curves diverge shortly after the first significant merger but follow similar trends thereafter. At late times, ongoing unclustered star formation depresses both \TN{} and \SM{}, and they trend downwards towards and below the values they had before any merger activity took place. The modifications made to the \ics{} affect the properties of the interstellar medium, and by extension they markedly influence the formation and disruption of the \gc{} populations. We note that the lack of an explicit prescription to model the cold phase of the interstellar medium in \emosaics{} artificially prolongs the survival of low-mass clusters.

\subsection{Contribution of star clusters to the stellar halo at \z[0]}
\label{sec:Results:gc_z0_properties}
Dynamical mass loss `transfers' stars from clusters into the field, and a large fraction of the stars lost from clusters can end up in the galaxy's stellar halo. In principle, so-called second-generation stars formed in clusters are identifiable via their anomalous light-element abundances: they are enhanced in He, N, Na, and Al, and depleted in C and O; properties that are found rarely in halo field stars \citep[e.g.][]{schiavon_chemical_2017,koch_purveyors_2019}. This enables their contribution to the stellar halo to be estimated observationally. Having shown that both the formation and the disruption of star clusters is sensitive to the galaxy assembly history, in this section we examine its influence on the fraction of the stellar halo contributed by disrupted star clusters.

There are several techniques that may be used to identify those stellar particles comprising the stellar halo in simulated galaxies \citep[see e.g.][]{zolotov_dual_2009,font_cosmological_2011}. Here, we follow the approach taken by \citet{hughes_fefeh_2020} and \citet{reina-campos_mass_2020}. We start by considering all stellar particles within \kpc[50] of the galaxy centre at \z[0], and discard any particles satisfying one or both of the following criteria:
\begin{enumerate}
    \item The azimuthal component of the angular momentum perpendicular to the galaxy disc, \Jz{}, is comparable to or greater than the angular momentum of a corotating circular orbit with similar orbital energy, \Jcirc{}, i.e. $\Jz{} \, /\, \Jcirc{} \geq 0.5$. We assume such stars comprise the rotationally supported disc \citep{sales_origin_2012}.
    \item The particle is within the galaxy's stellar half-mass radius, since here any dispersion-supported stars would be classified as comprising the galactic bulge.
\end{enumerate}
The choice of the outer radius (\kpc[50]) is somewhat arbitrary; however, it facilitates comparisons with observational analyses of the Milky Way and other \Lstar{} galaxies \citep[e.g.][]{koch_purveyors_2019}.

\begin{figure}%
    \centering%
	\includegraphics[width=\columnwidth]{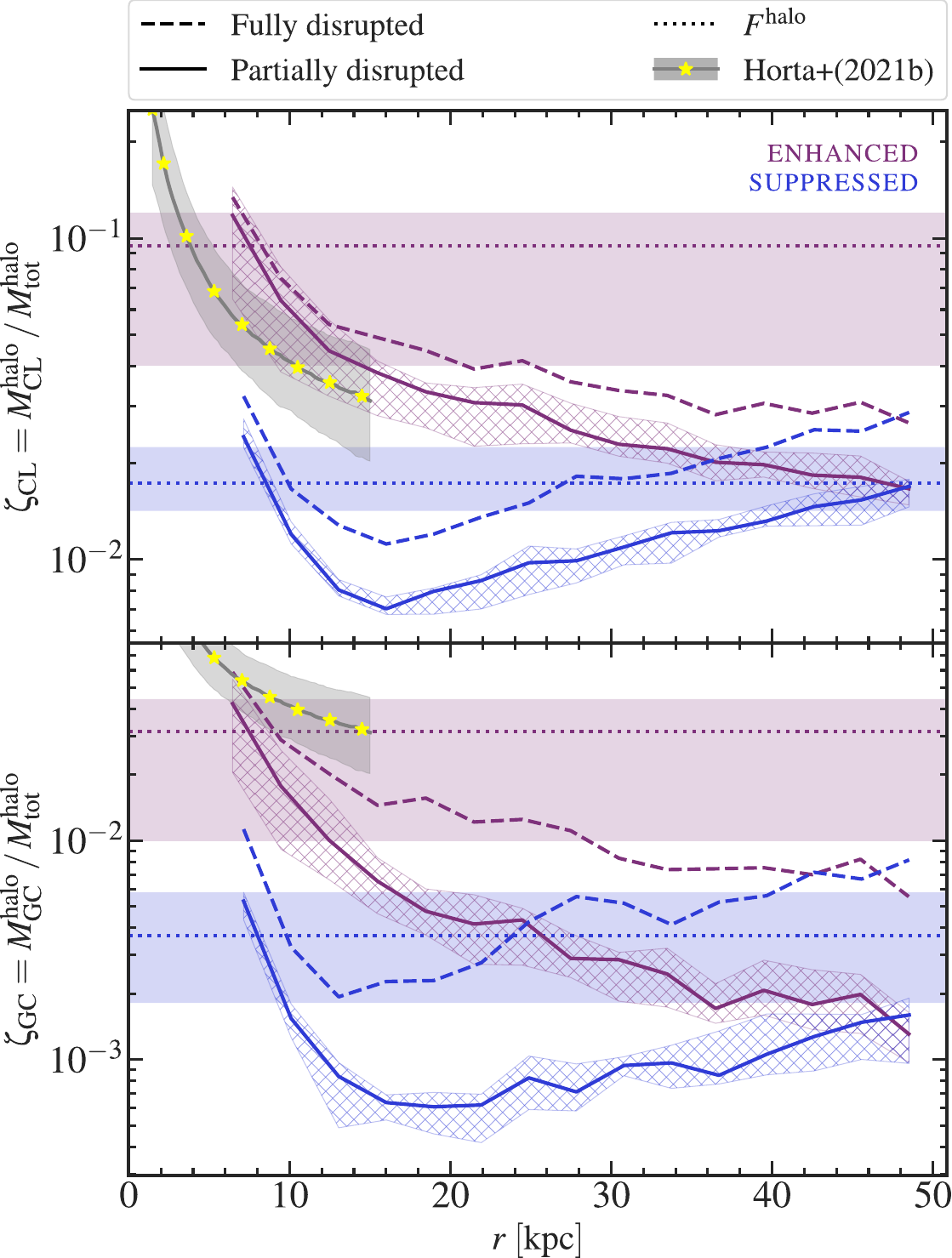}\\
	\caption{The fraction of the stellar mass in halo field stars that originated in star clusters as a function of the distance, \textit{r}, from the centre of the galaxy. Solid curves and hatched regions show the median mass lost to the field from the clusters and the associated \percent{68} scatter. Dashed curves and shaded regions show the same quantities assuming that all clusters surviving until \z[0] also fully disrupt, setting a strict upper limit on the fractional contribution of clusters to the stellar halo. The scatter due to differing choices of random number seed within the same family of simulations is similar to the hatched regions, and is not plotted for clarity.     As the mass fractions from the \Organic{} case are similar to those from the \Enhanced{} case, they are not shown. The curves marked with stars, and the associated shaded regions, show the total contribution of disrupted clusters to the Galactic stellar halo under the `minimal scenario' assumption estimated by \citet{horta_contribution_2021}. Upper panel: the mass fractions from all star clusters, $\zeta_{\rm CL}$, that formed at any time during the assembly of the galaxies. Bottom panel: the mass fractions from the subset of star clusters that formed as \gc{}s, i.e. with $M^{\rm birth}_{\rm cl}{>}10^5\Msun$, $\zeta_{\rm GC}$. For comparison, in both panels we overlay as dotted curves and shaded regions the median and \percent{68} scatter of \Fhalo{}; see \eqnref{eq:Fhalo}.
	}%
	\label{fig:Results:radial_fhalo}%
	\vspace{-10pt}%
\end{figure}

\figref{fig:Results:radial_fhalo} shows, as a function of galactocentric radius, the fraction of the mass of halo field stars that is contributed by stars born in
\begin{enumerate*}
    \item star clusters of any mass ($\zeta_{\rm CL}$, upper panel), and
    \item \gc{}s i.e. $M^{\rm birth}_{\rm cl}{>}\Msun[10^5]$ ($\zeta_{\rm GC}$, lower panel).
\end{enumerate*}
Solid curves show the fiducial outcome of the simulations, while dashed curves show the mass fraction that would be obtained if all clusters (upper panel) or \gc{}s (lower panel) were completely disrupted. As the mass fractions of the \Organic{} and \Enhanced{} cases are very similar, we do not plot the curves for the \Organic{} galaxy.

The dynamical disruption of star clusters is strongest close to the centre of the galaxy where the tidal field is stronger, and becomes less important at larger radii. This indicates that following major mergers, disrupted clusters originating from the accreted galaxy are dispersed into the halo of the descendent alongside the stars from disrupted clusters that formed \textit{in situ}. Consequently, in the cases with rich merger histories as much as \percent{15} of the mass in the stellar halo within \kpc[10] of the galaxy originated in star clusters (see \figref{fig:Results:radial_fhalo}, upper panel). This is consistent with the upper limit of \percent{11} derived from observations of the Milky Way while attempting to correct for survey selection effects \citep{koch_purveyors_2019}. At the outer extremity of the halo, i.e. \kpc[50], field halo stars that were born in star clusters account for only \percent{1.5-2} of the stellar halo mass. 

In the \Suppressed{} case, at all radii the fraction of field halo stars born in clusters is low, at less than \percent{2}. The different assembly histories therefore foster significant differences in the fraction of the stellar halo that is composed of stars disrupted from clusters. In the cases with rich merger histories the merging galaxies are sufficiently massive that \CFE{} and \Mcstar{} are already elevated in those systems prior to their merger with the target galaxies. The high gas densities reached during the mergers enhance the disruption of newly formed star clusters because of tidal shocks induced by the ambient gas surrounding the natal environment. This process, known as the `cruel cradle effect' \citep{kruijssen_dynamical_2012}, disrupts almost as many clusters as are born. Both aspects are crucial in the accumulation of large contributions of stars in the stellar halo originating from disrupted clusters.

In the rich merger histories, the elevated \CFE{} and \Mcstar{}, in combination with the cruel cradle effect, contribute to a higher total fraction of stars in the galaxy born in star clusters relative to the \Suppressed{} case. In that scenario, most halo stars are accreted in minor mergers that do not promote cluster formation, and within the central \kpc[10] star clusters contribute only \percent{3} of the stellar halo mass. Moreover, $\zeta_{\rm CL}$ decreases more rapidly with radius than in the cases with rich merger histories, reaching a minimum of much less than \percent{1} at $r \simeq \kpc[15]$. This pronounced feature arises from a peak in the population of field stars near the disc radius, which suggests that many halo stars formed within the disc and subsequently migrated out \citep[see e.g.][]{font_cosmological_2011}. At larger radii $\zeta_{\rm CL}$ steadily increases as a function of radius until $r=\kpc[50]$, where the value is similar to the \Enhanced{} case. This is consistent with observational evidence from the Milky Way suggesting that approximately \percent{70} of its stellar halo formed from the accretion of tidally disrupted low-mass galaxies \citep[e.g.][]{deason_progenitors_2015,belokurov_co-formation_2018,helmi_merger_2018,conroy_resolving_2019,mackereth_weighing_2020}. We defer a detailed analysis of the \textit{in-situ} and \textit{ex-situ} component of cluster formation to a future study.

Only a small fraction of the mass of stars born in clusters is contributed by \gc{}s, i.e. clusters whose initial masses are greater than $\Msun[10^5]$. Therefore, stars disrupted from \gc{}s comprise a much smaller fraction of the stellar halo than less massive clusters. In the \Suppressed{} case, \gc{}s contribute less than \percent{0.8} of the stellar halo within the central \kpc[10], and reach a minimum of \percent{0.04} at $r=\kpc[15]$ (see \figref{fig:Results:radial_fhalo}, bottom panel). In contrast, in the cases with rich merger histories up to \percent{5} of the central \kpc[10]  of the stellar halo originated in \gc{}s. In all three cases, the radial profile of $\zeta_{\rm GC}$ broadly follows a similar trend to $\zeta_{\rm CL}$: in the cases with rich merger histories the mass fraction decreases with radius; in the \Suppressed{} case we find an opposite, rising trend beyond \kpc[15]. At \kpc[50], stars from \gc{}s account for \percent{0.1} of the total stellar halo mass. In the inner halo the results from the cases with rich merger histories are broadly consistent with the `minimal scenario' upper limit of $\zeta_{\rm GC}$ \citep{schiavon_chemical_2017} presented by \citet{horta_contribution_2021} based on observations of the Galactic stellar halo. In the cases with rich merger histories, $\zeta_{\rm GC}$ and $\zeta_{\rm CL}$ bracket the observational estimate because the \citeauthor{horta_contribution_2021} sample contains stars originating from clusters with initial masses lower than \Msun[10^5], which we do not regard as \gc{}s. Our results are also consistent with other observational upper limits based on careful modelling of the outer stellar halo of the Milky Way \citep[e.g.][]{martell_chemical_2016,koch_purveyors_2019}. The inferred Galactic accretion history may thus be represented best by an intermediate assembly history between the \Suppressed{} and \Organic{} cases \citep[e.g.][]{deason_broken_2013,gallart_uncovering_2019,kruijssen_kraken_2020,naidu_reconstructing_2021}.

In \emosaics{}, the disruption of star clusters via shocks depends strongly on the density of the nearby star-forming gas \citep{pfeffer_e-mosaics_2018}. As the \Eagle{} simulations do not include an explicit treatment of the multiphase interstellar medium, they impose an artificial pressure floor that smooths density fluctuations in the interstellar medium and causes the tidal forces it generates to be too weak. Consequently, \emosaics{} underestimates the disruption rate of star clusters \citep{pfeffer_e-mosaics_2018}. We therefore interpret the mass lost from star clusters via dynamical disruption (the solid curves in both panels of \figref{fig:Results:radial_fhalo}) as a lower limit to the predicted value. We estimate an upper limit by including the mass of surviving star clusters in the mass in halo field stars (see the dashed curves in both panels in \figref{fig:Results:radial_fhalo}). This effectively assumes an extreme scenario in which all clusters are fully disrupted by \z[0]. We typically find that this upper limit corresponds to \percent{30} more halo stellar mass originating in star clusters compared to the fiducial outcomes of the simulations.

Previous observational and theoretical studies have attempted to characterize the total fractional contribution of stars born in clusters to the build-up of the stellar halo \citep[e.g.][]{martell_light-element_2010,martell_chemical_2016,schiavon_chemical_2017,koch_purveyors_2019,hanke_purveyors_2020,reina-campos_mass_2020,hughes_fefeh_2020,horta_contribution_2021}. To compare with these results we calculate the radially integrated contribution of the dynamically disrupted mass of star clusters and \gc{}s to the total mass of the stellar halo, $\Fhalo{}\!\left(z\right)$. We follow \citet{reina-campos_mass_2020} and define this quantity with the expression
\begin{equation}
    \label{eq:Fhalo}
    \Fhalo{}\!\left(z\right) = \frac{1}{\Mstarh{}\!\left(z\right)} \sum_i \Mclusti[z_{\rm birth}]{} f_{\ast,\, i}\!\left(z\right) - \Mclusti[z]{}\,,
\end{equation}
where $\Mstarh{}\!\left(z\right)$ is the total mass in halo field stars; \Mclusti[{\z{}}]{} is the total mass in clusters associated with star particle \textit{i}; and ${f_{\ast,\, i}\!\left(z\right) = \Mstari{}\!\left(z\right)\, /\, \Mstari{}\!\left(z_{\rm birth}\right)}$ is the fraction of the mass lost by star particle \textit{i} due to stellar evolution from birth until redshift, \z{}. We calculate $\Fhalo{}\!\left(\z[0]\right)$ for the galaxy in each assembly history and plot in \figref{fig:Results:radial_fhalo} the median and \percent{68} scatter as horizontal dotted curves and shaded regions, respectively. As \Fhalo{} is a cluster mass-weighted quantity, it primarily reflects the composition of the inner stellar halo where the contribution from stars born in clusters is greater. As such, its value is consistent with the inner bins of the radius-dependent fractions (solid curves), and generally it is systematically higher than the fractional contributions found in the outer haloes of the simulated galaxies. The values of \Fhalo{} we obtain are consistent with the observational upper limits on the fractional halo contribution from star clusters \citep{koch_purveyors_2019,horta_contribution_2021}; however, in general the fractional contribution of stars from \gc{}s in the simulations is much lower than this.

\section{Summary and Discussion}
\label{sec:Sum_Disc}
The influence of galaxy mergers on the formation of \gc{}s has been well studied. The ambition of this work is to explore, using novel controlled numerical experiments, how major mergers influence star cluster formation, disruption, and their contribution to the build-up of the stellar halo in an \Lstar{} galaxy. To do this we created a new suite of zoom-in simulations of the galaxy as part of the \emosaics{} project, using the \Eagle{} hydrodynamics scheme coupled with the \mosaics{} star cluster subgrid model. To isolate the effects of the galaxy mergers from other confounding influences, we use the genetic modification technique to change the assembly history of the galaxy while preserving its large-scale galactic environment. Previous work has often used large statistical samples of galaxies from diverse environments to explore the influence of merger history on galaxy properties; however, controlling for multiple variables in this type of analysis is difficult and can introduce correlated uncertainties. 

We have adopted a reference simulation, denoted as the \Organic{} case, in which an \Lstar{} galaxy experiences significant mergers at \z[1.7] and 0.8. Using this as a basis for comparison, we generated two distinct genetically modified \ics{} targeting the second major merger: one in which the mass ratio of the merging galaxies is doubled (\Enhanced{} case), and the other in which the merger does not take place at all (\Suppressed{} case). Our modifications alter the galaxy's assembly history while leaving its \z[0] halo mass unchanged. We allow the baryon properties to vary freely (see \secref{sec:Results:Galaxy_properties}). We accounted for stochastic variability in the galaxy properties by simulating the evolution of each set of \ics{} nine times using different random number seeds. The properties of the galaxy at a given time are therefore the median value of the property measured for all nine simulations.

Our findings can be summarized as follows:
\begin{enumerate}
    \item The assembly histories of galaxies imprint differences in the growth of \Mgc{} that manifest at \z[0] (see \figref{fig:Results:galaxy_properties}). The increased significance of the mergers in the \Enhanced{} case increase the rate of clustered star formation, leading to higher \Mgc{} at the present day compared to the \Organic{} case. In comparison, removing the target merger entirely suppresses the formation of \gc{}s and leads to lower \Mgc{}. The stratification of \Mgc{} in response to the assembly history begins to emerge as early as \tlb[{\Gyr[4]}].

    \item The feedback from the \SMBH{} plays an important role moderating the availability of star-forming gas in the central regions. This affects the disruption of star clusters throughout the lifetime of a galaxy. In the \Enhanced{} case the feedback generated by the \SMBH{} after the first significant merger contributes to the depletion of the cold gas reservoir, which is eroded further during the second merger (see \figref{fig:Results:galaxy_mgassf_mbh}). In the other families of simulations the response of the gas reservoir is more subdued, which illustrates that the significance of major mergers helps to establish the properties of the galaxy at \z[0]. In the cases where most of the gas reservoir has been removed, the resulting gas-poor environment can inhibit the disruption of star clusters in the \mosaics{} model.

    \item The \gc{} populations of present-day \Lstar{} galaxies are affected significantly by the major merger history of their host partly because it affects the evolution of \Mcstar{}. The \Enhanced{} case has more massive and fewer low-mass clusters compared to the \Organic{} case because conditions in the \Enhanced{} scenario preferentially favour massive cluster formation (see \figref{fig:Results:galaxy_z0_cluster_mass_function}). Conversely, in the \Suppressed{} case there are significantly more clusters with masses below \Msun[{10^4}] because of underdisruption driven by the more slowly varying tidal field and limitations in modelling the interstellar medium in \Eagle{}. The \Suppressed{} case is also in reasonable agreement with measurements of the abundance of star clusters in M31, and we expect that this would be improved by including an explicit model of the cold gas phase of the interstellar medium.
    
    \item Major mergers increase both the formation and the disruption of \gc{}s (see \figref{fig:Results:galaxy_sfr_gfr}). In our controlled experiments of this \Lstar{} galaxy, the first major merger stimulates the formation of over \percent{80} of the maximum mass in \gc{}s. In this example, major mergers at late times also enhance the rates of \gc{} formation and disruption, however the magnitude of these changes is smaller because the \gc{} system responds strongly to the availability, or the lack thereof, of supplies of cold star-forming gas in the central region of the galaxy at that time. We stress that the precise influence of mergers on the \gc{} population will be sensitive to the details of how the galaxy assembles.

    \item Major mergers pressurize the star-forming gas, which increases \CFE{} and \Mcstar{} and boosts the formation of \gc{}s (see \figref{fig:Results:mosaics_model_properties}). The magnitude of the enhancement in \gc{} formation rates depends on the availability of star-forming gas. For example, at late times in the \Enhanced{} case most of the gas has been expelled far beyond the central region of the galaxy or heated until it is no longer star-forming. This impedes new cluster formation despite elevated \CFE{} and \Mcstar{} in the remaining centrally located gas. In contrast, the galaxy in the \Organic{} case retains a larger star-forming gas reservoir at late times. Consequently, the boost in the \gc{} formation rate after the second major merger is larger than the \Enhanced{} case, even though \CFE{} and \Mcstar{} are similar.
    
    \item The specific mass, \SM{}, may be used to discriminate between the assembly histories of similar galaxies because it becomes stratified according to the merger history, while the specific frequency, \TN{}, does not (see \figref{fig:Results:galaxy_TN_SM}). The different behaviours of both metrics arise because they are sensitive to different properties of the natal environment. While both depend on the evolution of \CFE{}, \SM{} is particularly sensitive to the propensity for massive cluster formation over the galaxy's lifetime, and thus also depends on the evolution of \Mcstar{}. During the early peak cluster formation epoch \Mcstar{} stratifies according to assembly history, and these differences are enhanced slightly during the second major merger. This pattern remains embedded in the subsequent evolution of \SM{} until \z[0], where in the \Enhanced{} case it is twice as large as the \Suppressed{} case.

    \item Major mergers stimulate cluster disruption, affecting the final composition of the stellar halo. In galaxies with rich merger histories, disrupted star clusters contribute at least \percent{7} of the mass in the stellar halo, and at least \percent{3} of field stars were formed in \gc{}s (see \figref{fig:Results:radial_fhalo}). In contrast, the fractional contribution of clusters to the mass in field stars is \percent{75} lower in the \Suppressed{} case, with star clusters and \gc{}s contributing only \percent{1.7} and \percent{0.4} of the mass in the field halo, respectively. The radial profile of $\zeta_{\rm CL}$ is similar in the cases with rich merger histories, contrary to expectations that the fractional contribution might be elevated in the \Enhanced{} case. The similarity between the two assembly histories results from feedback during the first merger in the \Enhanced{} case that causes the galactic environment to be gas-poor during the final merger. This inhibits further cluster formation and disruption processes, such as shock heating \citep{pfeffer_e-mosaics_2018}. The value of \Fhalo{} that we calculate for the \Enhanced{} case could therefore be underestimated compared to a similar galaxy in which the stellar mass ratio of the first significant merger is smaller.
\end{enumerate}

Major mergers precipitate dramatic changes in the properties of the baryon component. In all three families of simulations the first significant merger drives large quantities of gas towards the central regions of the galaxies. This triggers bursts of star formation (see \figrefs{fig:Results:galaxy_mgassf_mbh}{fig:Results:galaxy_sfr_gfr}), and elevates the pressure (and density) of the star-forming gas in the coalescing galaxies by factors of $10^3-10^4$ above the ambient pressure in the interstellar medium prior to the merger. These conditions promote the highly efficient formation of massive clusters during the most intense period of star formation in the galaxy's history. Consequently, the first major merger inaugurates an early epoch of \gc{} formation during which more than \percent{80} of the maximum mass in \gc{}s forms. The duration of this epoch is influenced by feedback from star formation activity and the growth of the \SMBH{}. In the \Enhanced{} case these processes curtail the formation epoch after \Gyr[2]; however, in the other families of simulations it is sustained for an additional \Gyr[{0.5-1}] because the feedback is less intense. Consequently, less gas is expelled before stars and clusters can form. Over time the conditions that are favourable for \gc{} formation subside, reviving only briefly in the cases that experience a second significant merger shortly after \tlb[{\Gyr[8]}] (\z[1]). This second period of \gc{} formation contributes negligibly to the total mass in clusters in the galaxies; however, the clusters that form capture information about this late phase of galaxy evolution \Gyr[{1-2}] after the previous cluster formation epoch ended.

Compared to the first merger, the target merger exerts a weaker influence over many properties of the galaxy; however, it plays an important role stimulating the disruption of \gc{}s at late times. This is reflected in the composition of the stellar haloes at \z[0] in the galaxies with rich merger histories, which have the largest fractions of stars stripped from star clusters and \gc{}s. In these galaxies, at least \percent{9} and \percent{3}, respectively, of the total mass in field stars within \kpc[50] is derived from these sources (see \figref{fig:Results:radial_fhalo}). Most stars originating from disrupted clusters are concentrated within \kpc[{10-15}] of the galaxy where the tidal interactions with the host are strongest. This is consistent with the observational upper limit presented by \citet{koch_purveyors_2019} that up to \percent{11} of the Milky Way stellar halo comprises stars stripped from star clusters, and a value of \percent{4.2} at $r=\kpc[10]$ reported by \citet{horta_contribution_2021}. Our results are also consistent with observational evidence that suggests most of the stellar halo of the Milky Way originated from the Gaia--Enceladus merger event. At larger radii $\left(\kpc[{r > 40}]\right)$, the effects of the major mergers become less important and the fractions of halo stars contributed by star clusters in each assembly history are the same.

Somewhat counterintuitively, the significance of the target merger does not strongly affect the final composition of the stellar halo population, despite its importance for encouraging the disruption of \gc{}s. We can see its effects most clearly by comparing the \Enhanced{} case with the galaxy in the \Suppressed{} case. In the latter, $\zeta_{\rm CL}$ and $\zeta_{\rm GC}$ are at least \percent{1.7} and \percent{0.4}, respectively, both of which are more than \percent{70} lower than in the rich assembly histories. This result is in very good agreement with \citet{reina-campos_mass_2020}, who found that \percent{2.3^{+0.7}_{-0.4}} and \percent{0.3^{+0.2}_{-0.1}} of the stellar haloes in Milky Way mass galaxies from the \emosaics{} simulation suite are contributed by star clusters and \gc{}s, respectively. Our results suggest that most of the galaxies in their sample did not experience significant mergers after \z[1].

The total fractional contribution of disrupted star clusters to the stellar halo, \Fhalo{}, depends on the assembly history of the galaxy. Galaxies that undergo two significant mergers during their assembly have values of \Fhalo{} at \z[0] that are approximately eight times higher than galaxies in which the second merger is suppressed. It is likely that the values we obtain are somewhat underestimated because the \Eagle{} hydrodynamics scheme in our simulations does not explicitly model the cold phase of the interstellar medium, which is responsible for most of the disruption of clusters \citep{elmegreen_disruption_2010,kruijssen_modelling_2011,bastian_stellar_2012,miholics_tight_2017,reina-campos_introducing_2022}. This partly explains the increased survival probability of low-mass clusters in the \Suppressed{} case compared with those in the other simulations (see \figref{fig:Results:galaxy_z0_cluster_mass_function}). We account for this in our assessment of the build-up of the stellar halo by considering the extreme scenario that all clusters are disrupted prior to \z[0]. This sets an upper limit on the maximum contribution of clusters to the mass in the stellar halo that we find is consistent with similar observational estimates. Future studies of \gc{} formation and disruption in such systems would benefit from including an explicit model of the multiphase interstellar medium.

The genetic modification technique that we employ is a powerful tool to disentangle the contributions of various competing and sometimes degenerate properties to the evolution and assembly of galaxies. However, this approach cannot induce specific changes to a galaxy's assembly history in perfect isolation; other minimal changes must be introduced to ensure that the desired constraints at \z[0] are met. While our modifications of the \ics{} of the \Organic{} galaxy successfully changed the significance of the target merger at \tlb[{\Gyr[8]}] (\z[1]), they also affected the timing of the merger, and increased the significance of the first merger in the \Enhanced{} case. As the \gc{} population is strongly affected by the first major merger, an interesting and complementary extension to this study would be to directly modify the first merger while leaving the second merger unchanged. This will reduce the importance of the other minimal changes to the \ics{}. We also note that our conclusions are determined from simulations of one galaxy, targeting an individual merger. Thus, there are limits to how broadly we can generalize our conclusions from this set of simulations alone.

\section*{Acknowledgements}
We thank the reviewer, Oleg Gnedin, for his comments that improved the quality of the manuscript.
ON is supported by grant 2020/39/B/ST9/03494 from the Polish National Science Centre, and was supported by the Royal Society during part of this study. RAC was supported by a Royal Society University Research Fellowship during part of this study.
JP was supported by the Australian government through the Australian Research Council's Discovery Projects funding scheme (DP220101863).
JMDK gratefully acknowledges funding from the European Research Council (ERC) under the European Union's Horizon 2020 research and innovation programme via the ERC Starting Grant MUSTANG (grant agreement number 714907). COOL Research DAO \citep{chevance_cool_2025} is a Decentralised Autonomous Organisation supporting research in astrophysics aimed at uncovering our cosmic origins.
AP and JD received funding from the European Union’s Horizon 2020 research and innovation programme under grant agreement No.~818085 GMGalaxies.
This work used the Prospero high performance computing facility at Liverpool John Moores University (LJMU).

\textit{Software}: This work made use of \astropy{} \citep{the_astropy_collaboration_astropy_2013,the_astropy_collaboration_astropy_2018}, \matplotlib{} \citep{hunter_matplotlib_2007}, \numpy{} \citep{walt_numpy_2011,harris_array_2020}, \pynbody{} \citep{pontzen_pynbody_2013}, \python{} \citep{van_rossum_python_2009}, \scipy{} \citep{jones_scipy_2011,virtanen_scipy_2020}, and \tangos{} \citep{pontzen_tangos_2018}. This research also made use of the NASA Astrophysics Data System~(\url{http://adsabs.harvard.edu/}) and the arXiv e-print service (\url{http://arxiv.org/}). We thank their developers for maintaining them and making them freely available.

\section*{Data Availability}
A repository of reduced data and scripts to produce the figures in this work is available on GitHub\footnote{Supporting Information: \github{Musical-Neutron/gc-gmics}} and archived in Zenodo \citep{newton_formation_2025}.



\bibliographystyle{mnras}
\bibliography{archive} 


\bsp	
\label{lastpage}
\end{document}